%% file: nc_post_accept1_arxiv.tex
\documentclass[reprint,amsmath,amssymb, aps,pre,floatfix,superscriptaddress]{revtex4-2}

\usepackage{graphicx} 
\usepackage{siunitx}
\usepackage[version=4]{mhchem}
\usepackage{comment}
\usepackage{xcolor}
\usepackage{tikz}
\usepackage{scalerel}
\usepackage{xifthen}
\usepackage{chngcntr}
\usepackage{xcolor}
\usepackage{bm}
\usepackage{physics}
\usepackage[utf8]{inputenc}
\usepackage[T1]{fontenc}
\usepackage{newfloat}
\usepackage{subcaption}
\usepackage{makecell}
\usepackage[page,header]{appendix}

\usetikzlibrary{svg.path}
\usepackage[colorlinks,allcolors=blue]{hyperref}

\definecolor{orcidlogocol}{HTML}{A6CE39}
\tikzset{
  orcidlogo/.pic={
    \fill[orcidlogocol] svg{M256,128c0,70.7-57.3,128-128,128C57.3,256,0,198.7,0,128C0,57.3,57.3,0,128,0C198.7,0,256,57.3,256,128z};
    \fill[white] svg{M86.3,186.2H70.9V79.1h15.4v48.4V186.2z}
 svg{M108.9,79.1h41.6c39.6,0,57,28.3,57,53.6c0,27.5-21.5,53.6-56.8,53.6h-41.8V79.1z M124.3,172.4h24.5c34.9,0,42.9-26.5,42.9-39.7c0-21.5-13.7-39.7-43.7-39.7h-23.7V172.4z}
 svg{M88.7,56.8c0,5.5-4.5,10.1-10.1,10.1c-5.6,0-10.1-4.6-10.1-10.1c0-5.6,4.5-10.1,10.1-10.1C84.2,46.7,88.7,51.3,88.7,56.8z};
  }
}

\newcommand\orcid[1]{\href{https://orcid.org/#1}{\mbox{\scalerel*{
\begin{tikzpicture}[yscale=-1,transform shape]
\pic{orcidlogo};
\end{tikzpicture}
}{|}}}}

\DeclareFloatingEnvironment[
  fileext=sfig,
  listname={List of Supplementary Figures},
  name=Fig.,
  placement=tbhp,
  within=none,
]{suppfigure}

\DeclareCaptionLabelFormat{custom}
{
  #1 S#2
}
\DeclareCaptionLabelSeparator{custom}{}
\DeclareCaptionFormat{custom}
{
    \small #1#2 #3
}

\begin{document}

\setlength{\unitlength}{1cm}

\newcommand{\nus}{Department of Mechanical Engineering, National University of Singapore, 117575, Singapore}
\newcommand{\hust}{State Key Laboratory of Coal Combustion, School of Energy and Power Engineering, Huazhong University of Science and Technology, Wuhan, Hubei, 430074, China}
\newcommand{\wut}{School of Naval Architecture, Ocean and Energy Power Engineering, Wuhan University of Technology, Wuhan, Hubei, 430063, PR China}
\newcommand{\padova}{Department of Industrial Engineering and CISAS ``G. Colombo'', University of Padova, Padova, 35122, Italy}

\input{sym_post_accept1}

\input{sym_math_post_accept1}

\title{\TITLEnew}

\author{Qianhong Yang~\orcid{0000-0002-4840-2039}}%
\thanks{These two authors contributed equally}
\affiliation{\nus}
\author{Maoqiang Jiang~\orcid{0000-0002-4371-2601}}
\thanks{These two authors contributed equally}
\affiliation{\wut}
\affiliation{\nus}
\author{Francesco Picano~\orcid{0000-0002-3943-8187}}
\affiliation{\padova}
\author{Lailai Zhu~\orcid{0000-0002-3443-0709}}%
\email{lailai\_zhu@nus.edu.sg}%
\affiliation{\nus}
\date{\today}%

\begin{abstract}
Active matter drives its constituent agents to move autonomously by harnessing free energy, leading to diverse emergent states with relevance to both biological processes and inanimate functionalities. Achieving maximum reconfigurability of active materials with minimal control remains a desirable yet challenging goal. Here, we employ large-scale, agent-resolved simulations to demonstrate that modulating the activity of a wet phoretic medium alone can govern its solid-liquid-gas phase transitions and, subsequently, laminar-turbulent transitions in fluid phases, thereby shaping its emergent pattern. These two progressively emerging transitions, hitherto unreported, bring us closer to perceiving the parallels between active matter and traditional matter. Our work reproduces and reconciles seemingly conflicting experimental observations on chemically active systems, presenting a unified landscape of phoretic collective dynamics. These findings enhance the understanding of long-range, many-body interactions among phoretic agents, offer new insights into their non-equilibrium collective behaviors, and provide potential guidelines for designing reconfigurable materials.
\end{abstract}

\maketitle

\section*{Introduction}
Active matter represents a class of material systems comprised of autonomous units capable of converting free energy into mechanical work~\cite{ramaswamy2010mechanics,gompper20202020}. The collective motion of these units, mediated by their interactions, can bring in fascinating self-organization, pattern formation, and coherent activities~\cite{bois2011pattern, vicsek2012collective, marchetti2013hydrodynamics}. Their emergence and maintenance are crucial in biological systems~\cite{karsenti2008self,needleman2017active}, inspiring the development of synthetic, autonomous agents as the foundation for reconfigurable, functional materials~\cite{aranson2013active, bishop2023active}. A key objective of designing such bio-inspired systems is to maximize their reconfigurability with minimal control. Prior research has demonstrated controlling activity to achieve either phase change in active matter~\cite{redner2013structure, cates2015motility} or laminar-turbulent transition in active fluids~\cite{wensink2012meso, alert2022active}, but not both. Here, we find that remarkably, tuning the activity of a phoretic medium alone can control its solid-liquid-gas phase transitions and subsequently, laminar-turbulent transitions in fluid phases. 
Through large-scale, agent-resolved simulations, we investigate suspensions of isotropic phoretic agents (IPAs) epitomized by active droplets~\cite{thutupalli2011swarming,banno2012ph, izri2014self,maass2016swimming,moerman2017solute,thutupalli2018flow,meredith2020predator,lohse2020physicochemical,de2020spontaneous,suda2021straight,dwivedi2022self,michelin2022self,matsuo2023sequentially,feng2023self} and camphor surfers~\cite{tomlinson1862ii}, explicitly resolving their many-body  hydrochemical interactions. 
Our dual consideration of long-range hydrodynamic and chemical interactions enables not only reproducing characteristic collective behaviours of IPAs observed in the lab, but also
reconciling seemingly divergent experimental observations---active crystallization or turbulence. The  unified landscape of phoretic collective dynamics is unattainable by resolving either the hydrodynamic or chemical interaction alone.

\begin{figure*}[tbh!]
\centering
\includegraphics[width=1\linewidth]{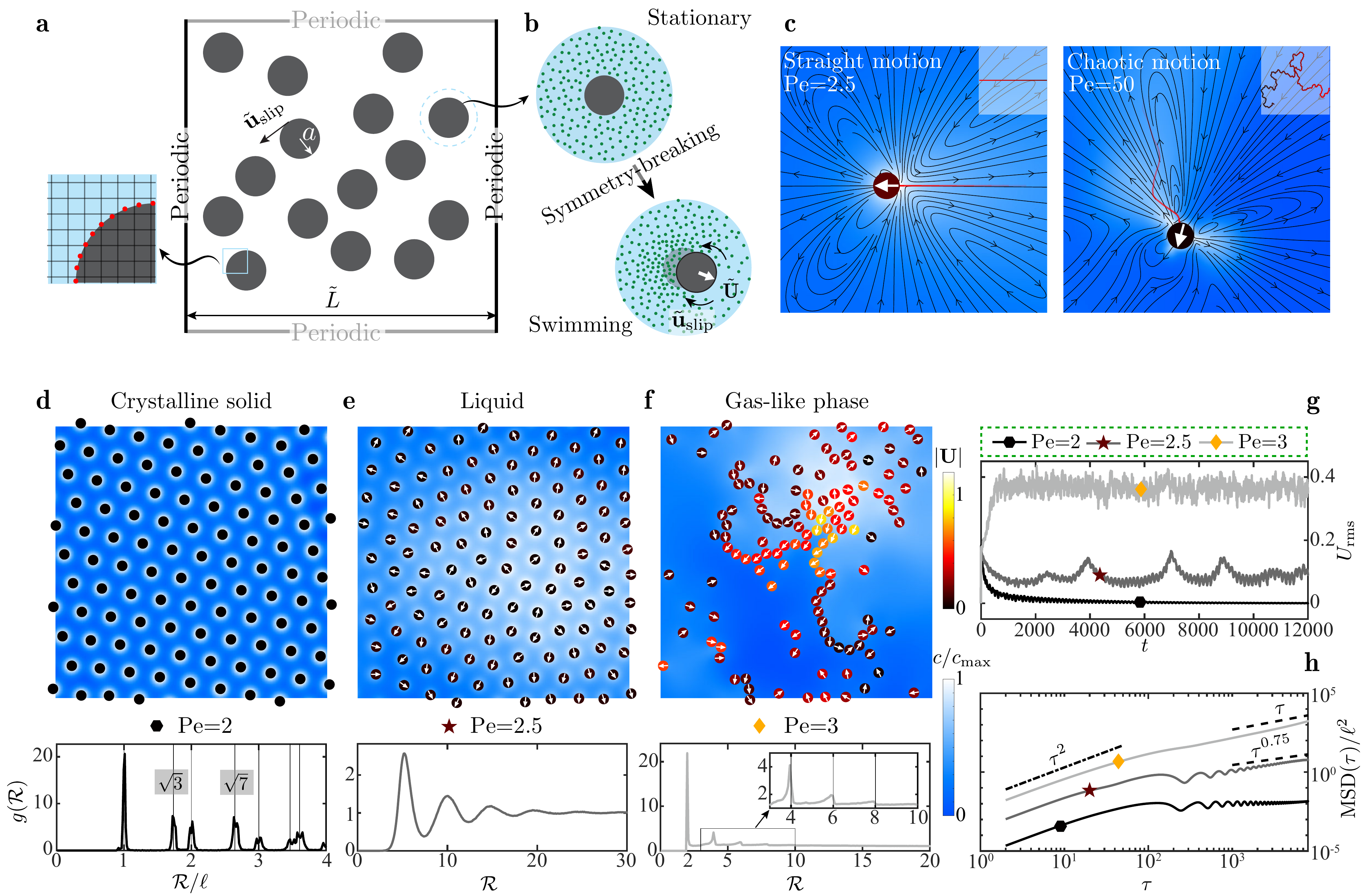}
\caption{
\textbf{Activity-induced phase transition.} \textbf{a}, $N$ circular phoretic disks of radius $a$ freely swimming in a periodic square domain of size $\tilde{L}$.
The notation $\tilde{}$~ represents dimensional variables throughout.
The side panel illustrates the numerical discretization (\SI). \textbf{b}, sketch of a single phoretic disk attaining a swimming velocity, $\tbU$, spontaneously via instability. Green dots indicate a chemical solute emitted by the disk, while $\tbuslip$ denotes its surface slip velocity induced by the chemical gradient.  \textbf{c}, a swimming disk of low (left) or high (right) activity---implied by the P\'eclet  number $\Pe$---follows a 
straight or chaotic trajectory, respectively. The colormap shows the scaled solute concentration $c/\cmax$ and the insets depict the trajectories. The streamlines surrounding the swimmer at $\Pe=2.5$  demonstrates its pusher-like dipolar signature, as previously identified numerically~\cite{morozov2019nonlinear} and experimentally~\cite{hokmabad2022spontaneously}.
\textbf{d}-\textbf{f}, disks with an area fraction $\phi=0.12$ self-organize into hexagonal solid ($\Pe=2$), liquid ($\Pe=2.5$), and gas-like ($\Pe=3$) phases depicted in a quarter of the domain. Disks are colored by their swimming speed $|\bU|$ and the arrow indicates the instantaneous direction of $\bU$. The second row displays the corresponding pair correlation function $g(\Rsep)$. Here, $\Rsep$ denotes the inter-disk distance and $\ell = \lp 2\pi \phi^{-1}/\sqrt{3} \rp^{1/2}$ represents the lattice constant.
\textbf{g} and \textbf{h}, root-mean-square (RMS) disk velocity $\avel$ versus time and the scaled mean square displacement (MSD) versus the time lag $\tau$, respectively, for varying $\Pe$.  
}
\label{fig:1}
\end{figure*}

We focus on a two-dimensional (2D) paradigmatic system of IPAs. An IPA, unlike Janus colloids~\cite{chen2011directed,walther2013janus}, acquires autonomous propulsion through instability, but otherwise in a stable stationary state. 
For instance, an active droplet undergoing uniform surface reaction exchanges solutes with the ambient, causing an isotropic solute distribution (Fig.~\ref{fig:1}{\textbf{b}}). A perturbation inducing a Marangoni interfacial flow advects the solute, amplifying itself. When the ratio of 
the destabilizing advection to the stabilizing solute diffusion---characterized by P\'eclet ($\Pe$) number, exceeds a threshold, instability emerges, driving the droplet to swim steadily (Fig.~\ref{fig:1}{\textbf{c}} left); increasing $\Pe$ may trigger its chaotic movement~\cite{morozov2019nonlinear,hu2019chaotic, hokmabad2021emergence,chen2021instabilities,kailasham2022dynamics,li2022swimming} (Fig.~\ref{fig:1}{\textbf{c}} right). Similarly, this mechanism allows camphor disks to swim continuously or intermittently~\cite{nakata2006self,suematsu2010mode}.

We consider $N$ overdamped disks of radius $a$ in a doubly-periodic square domain of size $\tilde{L}$ (Fig.~\ref{fig:1}{\textbf{a}}). Hereinafter, $\tilde{}$~ denotes dimensional variables.  The disks uniformly emit a chemical solute of molecular diffusivity $\mD$ at a constant rate $\mA>0$.
The solute distribution $\tc\left[ \tbr = \lp \tx,\ty \rp, \tt \right] $ causes a slip velocity $\tbuslip=\mM \lp \bI - \bn\bn \rp \cdot \tgrad \tc $, with $\mM$ the mobility coefficient and $\bn$ the outward normal at the disk surface. 
We vary the phoretic activity $\Pe =\mA \mM a /\mD^2$ and area fraction $\phi= \pi N /L^2$ ($L=\tilde{L}/a$) of disks to explore their collective dynamics, solving dimensionless physicochemical hydrodynamics involving the fluid velocity $\bu$ and pressure $p$, and solute concentration $c$, see \mat~and Supplementary Information (\SI).

\section*{Results}

\subsection*{Crystalline Solid, Liquid, and Gas-like Phases}

A disk suspension of $\phi=0.12$ exhibits $\Pe$-dependent collective patterns (Fig.~\ref{fig:1}{\textbf{d}-\textbf{f}}). At $\Pe=2$, disks undergo transient, disordered motion that decays in time. Hence, their root-mean-square (RMS) velocity $\avel=\sqrt{N^{-1}\sum^N_{k=1} \bU^2_k}$ ($\bU_k$ is the translational velocity of the $k$-th disk) eventually approaches zero, when they self-organize into a stationary hexagonal lattice resembling a 2D crystalline solid  (Fig.~\ref{fig:1}{\textbf{d}} and \movRef{1}). This resemblance is supported by the pair correlation function $g(\Rsep)$ versus the inter-disk distance $\Rsep$ (\SI)
depicted in Fig.~\ref{fig:1}{\textbf{d}}, which reveals the signature of a 2D hexagonal crystal with a lattice constant $\ell= \lp 2\pi\phi^{-1}/ \sqrt{3}  \rp^{1/2}$: $g(\Rsep)$ peaks around discrete inter-disk distances $\Rsep/\ell=1,\sqrt{3},2,\sqrt{7},...$. Importantly, the lattice constant $\ell \approx 5.5$ considerably larger than the disk diameter implies that this solid shares similarity with the Wigner crystal constituting electrons, as theoretically predicted~\cite{wigner1934interaction, wigner1938effects} and directly visualized in experiments very recently~\cite{li2021imaging}. Unlike the long-range Coulomb force causing the electronic Wigner crystallization, the chemorepulsion among phoretic disks~\cite{moerman2017solute,jin2017chemotaxis,michelin2022self} creates the active Wigner crystal.  This phenomenon agrees with the experimental observations on camphor surfers~\cite{soh2008dynamic} and active droplets~\cite{thutupalli2018flow}, as well as numerical predictions for the former~\cite{gouiller2021two}. Such active Wigner crystals are distinct from the  hexagonal closed-packed crystallization commonly reported in other active suspensions~\cite{bialke2012crystallization, palacci2013living,petroff2015fast,briand2016crystallization, singh2016universal,klamser2018thermodynamic, kichatov2021crystallization,tan2022odd}, where active units  tend to achieve the highest packing fraction akin to the atomic arrangement in graphene layers and some metals.

\begin{figure}[bh!]
\centering
\includegraphics[width=1\linewidth]{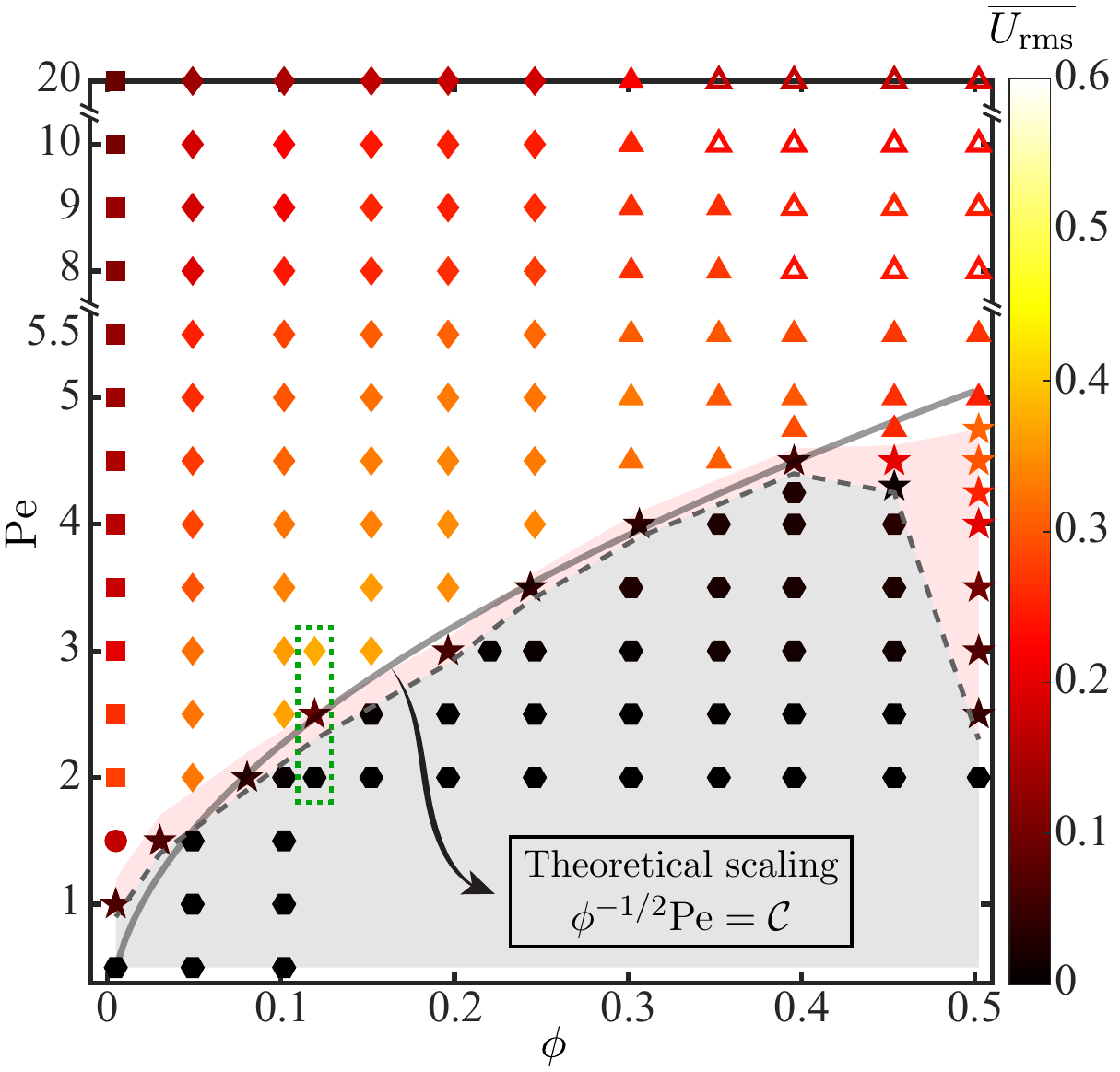}
\caption{\textbf{Phase diagram over the area fraction $\phi$ and activity $\Pe$ of phoretic disks.} The disks form crystalline solids (hexagon),  liquids (star), or gas-like phases consisting of four classes: no pattern (circle); single stable chain (square); dynamic chaining (diamond); and self-organized coherent flows (triangle). Symbols are colored by the time-averaged RMS velocity $\avelt$ of disks. The solid curve represents the theoretical prediction Eq.~[\ref{eq:scaling}] of the solid-liquid transition, where $\mC$ denotes a fitting parameter. The dashed box encloses the three phases showcased in Fig.~\ref{fig:1}.
}
\label{fig:2}
\end{figure}

Increasing the activity to $\Pe=3$, the disks freely swim like Brownian gas molecules (Fig.~\ref{fig:1}{\textbf{f}} and \movRef{2}), with randomly fluctuating speed $\avel$ (Fig.~\ref{fig:1}{\textbf{g}}).  After a transient period, this fluctuating motion exhibits normal diffusion, as evidenced by the mean square displacement (MSD) of disks (Fig.~\ref{fig:1}{\textbf{h}}). Dynamic arch-shaped chains of closely spaced disks also appear, similar to chains of active droplets observed experimentally~\cite{thutupalli2011swarming,thutupalli2018flow}. These chains form, collide, annihilate, and reform continuously (\movRef{2}). This dynamic chaining differs from the single stable chains observed at lower fraction $\phi$, which originate from a balance between inter-disk chemorepulsion and hydrodynamic attraction (\SI). Overall, this suspension reveals features of both a gas phase and chain formation, as statistically evinced in Fig.~\ref{fig:1}{\textbf{f}}: $g(\Rsep) \approx 1$ in most region of $\Rsep\geq 2$, characteristic of a homogeneous molecular distribution; it develops peaks with decreasing magnitudes approximately at even multiples of the disk radius, \ie $\Rsep \approx 2$, $4$, and $6$---that signify chain forming.

Between the solid and gas-like phases, a liquid phase  appears at $\Pe=2.5$ (Fig.~\ref{fig:1}{\textbf{e}} and \movRef{3}). The disks vibrate, move around, and slide over each other mimicking the behavior of liquid molecules. The $\avel$ displays a super-oscillatory time evolution  (Fig.~\ref{fig:1}{\textbf{g}}), and the $\text{MSD}\propto\tau^{0.75}$ exhibits subdiffusive dynamics (Fig.~\ref{fig:1}{\textbf{h}}).  This anomalous diffusion, unlike the normal diffusion of classical liquids, is typical of  crowded fluids containing suspended macro-molecules or colloids that act as obstacles~\cite{banks2005anomalous,golding2006physical}. The identified liquid phase is confirmed by the pair correlation function $g(\Rsep)$ depicted in Fig.~\ref{fig:1}{\textbf{e}}, where several oscillations at short distances $\Rsep$ attenuate with increasing $\Rsep$, indicating the diminished long-range order characteristic of liquids.

\begin{figure*}[tbh!]
\centering
\includegraphics[width=1\linewidth]{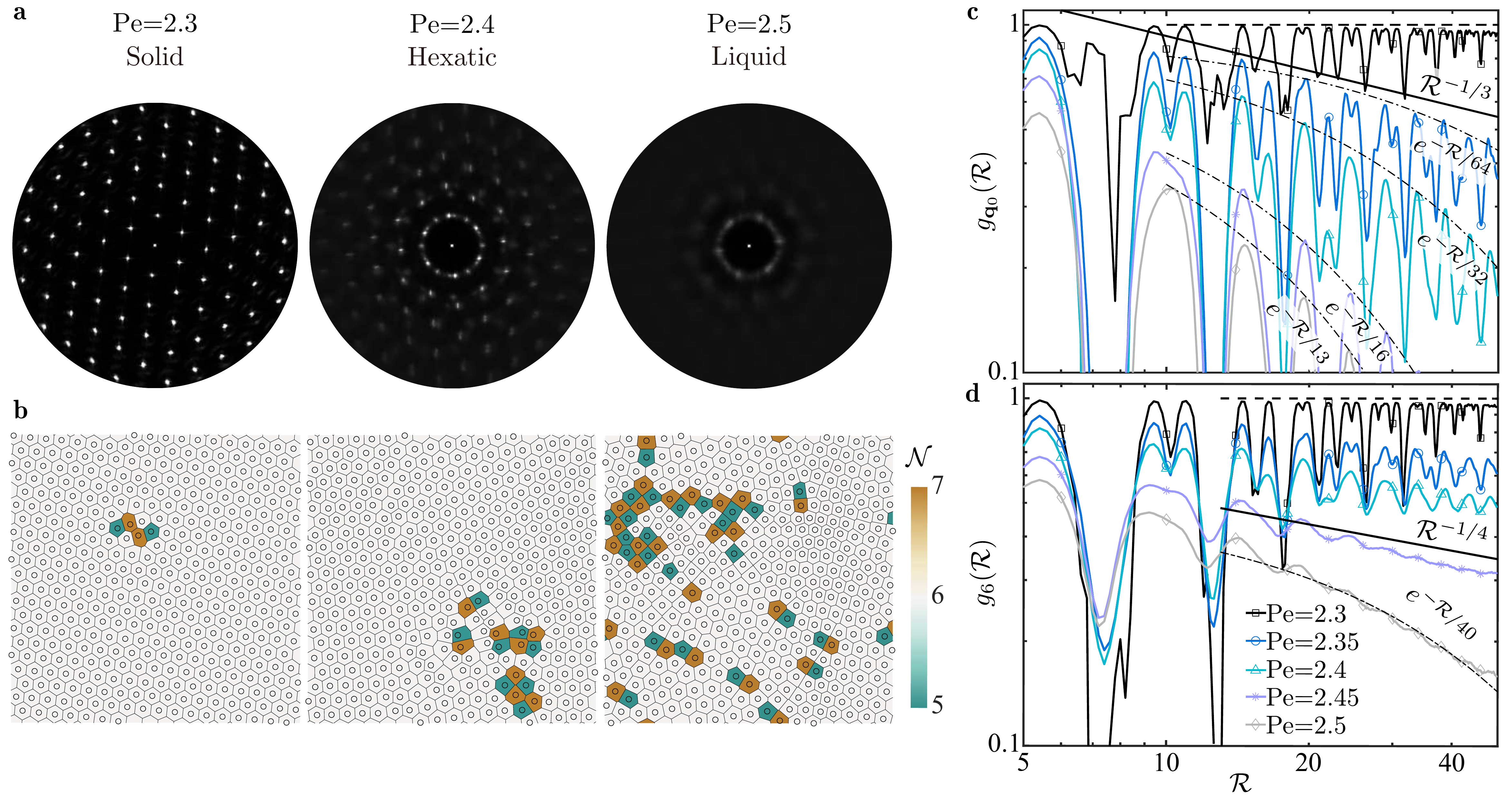}
\caption{
\textbf{Defect-mediated melting via a hexatic phase.}
\textbf{a}, structure factor $S(\bq)$ of a $\phi=0.12$ disk suspension at three $\Pe$ values, indicative of a growing effective temperature. \textbf{b}, Evolving topological defects associated with the three phases in \textbf{a}, showing a bound dislocation pair, two free dislocations, and coexisting dislocations and disclinations, consecutively. Disks with $\Nnb=5$ and $7$ neighbours are marked green and yellow, respectively.  
\textbf{c}, translational order correlation function $\gtcfR$ at $\Pe\in[2.3, 2.5]$; the horizontal dashed line indicates a constant, non-decaying  $\gtcfR$. \textbf{d}, similar to \textbf{c}, but for the orientational order correlation function $\gocfR$. }
\label{fig:3}
\end{figure*}

\subsection*{Phase Diagram and Theoretical Prediction of the Solid-Liquid Transition}

Displaying the time-averaged mean disk velocity $\avelt$ versus the fraction $\phi$ and activity $\Pe$ in Fig.~\ref{fig:2}, we reveal the $\phi-\Pe$ phase diagram of the suspension. Besides the solid-liquid-gas transitions, we divide the gas-like phase into four regimes: no pattern; single stable chain; dynamic chaining (Fig.~\ref{fig:1}{\textbf{f}}); and self-organized large-scale flows exhibiting instability, transition and active turbulence, which will be discussed later; the first two regimes are analyzed in \SI. For now, we focus on theoretically explaining the solid-liquid transition.

Upon increasing $\Pe$ to cross a threshold (dashed curve in Fig.~\ref{fig:2}), the suspension transitions from a quiescent hexagonal pattern to unsteady motion. This change is in reminiscent of the stationary-to-swimming instability of an IPA with sufficient activity $\Pe$. The theoretically predicted onset of instability for a single IPA~\cite{michelin2013spontaneous,hu2019chaotic,schnitzer2022weakly,zhu2023self} provides inspiration for a prediction that characterizes the collective instability of our suspension.

We consider the hexagonal tiling of disks as the unperturbed state. Following $\Pe=\mA \mM a /\mD^2$, which determines the instability of a single disk, we define $\Pec$ for the collective instability of many featuring $\phi$. From the dimensional advection-diffusion equation (\mat), we infer
\begin{align}\label{eq:pec}
    \Pec = \frac{\left[ \tbu 
    \right]\left[ \tgrad \tc \right]}{\mD \left[ \tgrad^2 \tc\right]},
\end{align}
using $\left[ \cdot \right]$ to represent the characteristic scales. As for a single disk,  $\left[\tbu \right]= \mA\mM/\mD$ remains. Further limited to the low-$\phi$ regime corresponding to a large inter-disk distance $\tell \gg a$ ($\ell \gg 1$), we regard the disks as points and thus approximate $|\tgrad \tc |$ at the disk center by its exact value $\mA/\mD$ at the disk boundary. At the midpoint of two neighbouring disks, $\tgrad\tc=\mathbf{0}$ due to the hexagonal symmetry (\SI). Hence, $|\tgrad \tc|$ decays from $\mA/\mD$ to zero within the  distance $\tell/2$, leading to $\left[\tgrad^2 \tc \right]=2\mA/\lp\mD\tell\rp$ and subsequently
\begin{align}\label{eq:scaling}
    \Pec = \lp \frac{\pi}{2\sqrt{3}} \rp^{1/2}  \phi^{-1/2} \Pe\propto  \phi^{-1/2}\Pe. 
\end{align}
Assuming that the instability occurs when $\phi^{-1/2}\Pe$ exceeds a constant $\mC$, we obtain, via fitting, the predicted threshold $\phi^{-1/2}\Pe =\mC$ (solid curve in Fig.~\ref{fig:2}) that matches the actual one reasonably
when $\phi \lessapprox 0.4$.

\begin{figure*}[tbh!]
\centering
\includegraphics[width=0.9\linewidth]{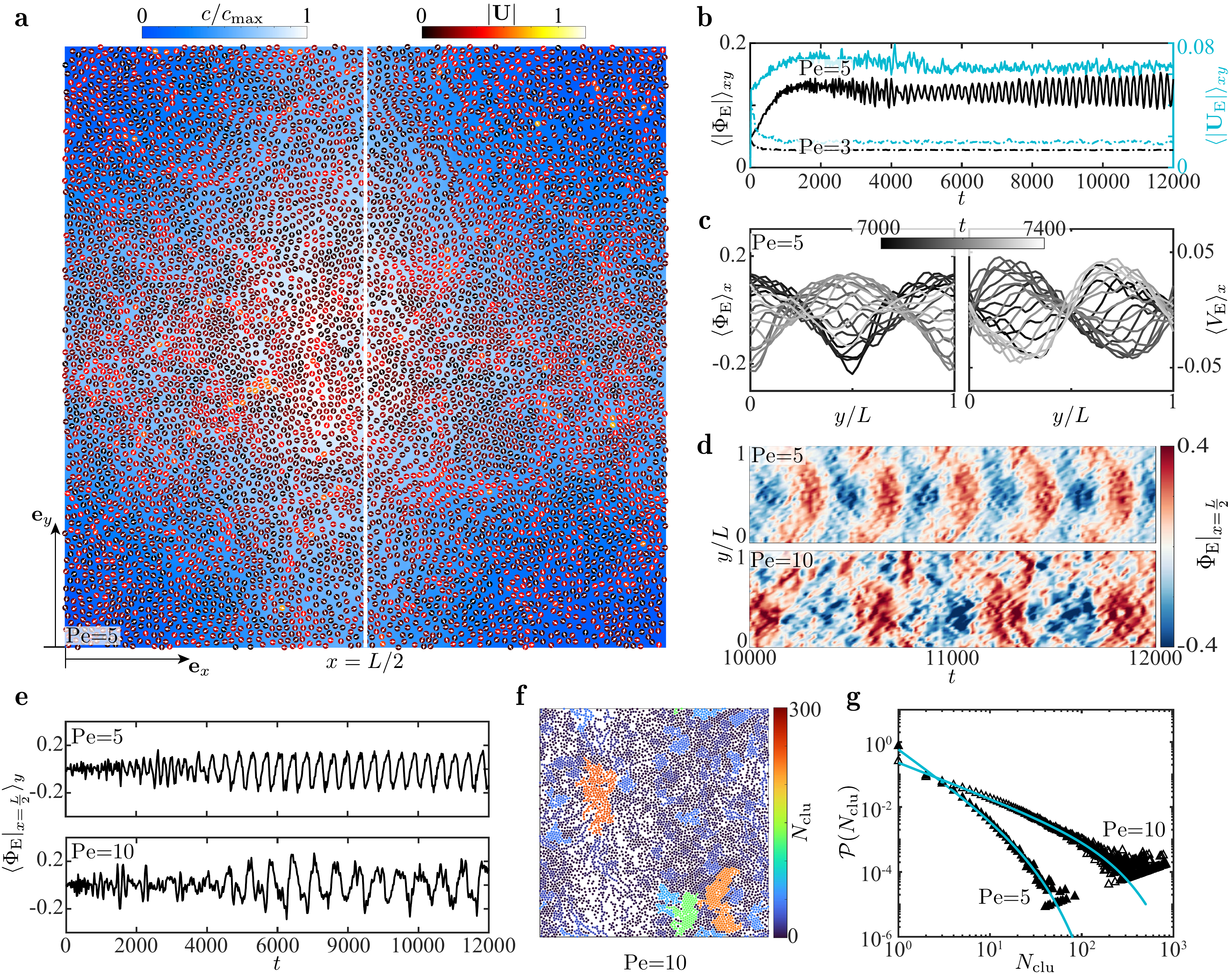}
\caption{\textbf{Instability and transition of the active fluid with $\boldsymbol{\phi=0.5}$.}
\textbf{a}, an oscillatory active flow self-organizes via instability at $\Pe=5$, when the dense population of disks shifts  periodically between the corners (this snapshot) and the center (\SI) of the domain. \textbf{b}, history of domain-averaged absolute pressure $\la |\phid| \ra_{xy}$ and speed $\la |\bUloc| \ra_{xy}$ of the active fluid implies the absence and emergence of a self-organized flow at $\Pe=3$ and $\Pe=5$, respectively. \textbf{c}, wave-like spatiotemporal evolution of $x$-averaged pressure $\la \phid \ra_{x}$ and velocity component $\la \Vloc \ra_{x}$ at $\Pe=5$. \textbf{d}, corrugated kymograph of $\phid|_{x=\frac{L}{2}}$ sampled along the median $x=L/2$ when $\Pe=5$ (upper) versus its disrupted counterpart for $\Pe=10$ (lower). \textbf{e}, history of $\la \phid|_{x=\frac{L}{2}} \ra_{y}$ depicts the evolution and saturation of disturbances mimicking those of their canonical hydrodynamic analogue.  
\textbf{f}, when $\Pe=10$, disks form clusters that breakdown the wave patterns. The clusters are characterized by the number $\Nswarm$ of their constituting disks. \textbf{g}, cluster size distribution function $\Pswm$, and its fitted curves following $\Pswm=\Ca \Nswarm^{-\Cb}\exp\lp -\Nswarm/\Nswmcut\rp$ defined in SI. 
}
\label{fig:4}
\end{figure*}

\subsection*{Two-Dimensional Melting via A Hexatic Phase}

Further tuning $\Pe$ as the activity-induced effective temperature, we scrutinize the solid-liquid transition at $\phi=0.12$ as a melting scenario. We identify an intermediate phase that is neither solid nor liquid. As indicated in Fig.~\ref{fig:3}{\textbf{a}}, the structure factor $S\lp \bq \rp$ (\SI) for $\Pe=2.3$ exhibits definite Bragg peaks with six-fold symmetry, revealing the formation of a crystalline solid. Raising the effective temperature to $\Pe=2.5$, the Bragg peaks are almost smeared out, leaving an isotropic ring pattern of $S\lp \bq \rp$ with insignificant orientational symmetry. This suggests that the suspension has melted, reaching a liquid state. At  $\Pe=2.4$, the translational order is lost while six-fold orientational symmetry preserves. This intermediate state corresponds to a hexatic phase between the solid and liquid, as described by the celebrated Kosterlitz, Thouless, Halperin, Nelson, and Young (KTHNY) theory~\cite{kosterlitz1973ordering,halperin1978theory,young1979melting}. This theory, originally built for equilibrium systems, has also been tested upon non-equilibrium counterparts of active agents~\cite{klamser2018thermodynamic,digregorio2018full,pasupalak2020hexatic,paliwal2020role,shi2023extreme}.

Within the KTHNY theory, melting of 2D crystals is mediated by progressive creation of topological defects, as illustrated in Fig.~\ref{fig:3}{\textbf{b}} for our case. When $\Pe=2.3$, we detect bound dislocation pairs. Increasing the activity, we observe the
dissociation of such pairs into free dislocations at $\Pe=2.4$ (\movRef{4}), and the unbinding of dislocations into isolated disclinations at $\Pe=2.5$ (\movRef{5}). The two successive processes drive the solid-to-hexatic and hexatic-to-liquid  transitions, respectively, which highlight the signature two-step storyline of the KTHNY framework.

\begin{figure*}[tbh!]
\centering
\includegraphics[width=1\linewidth]{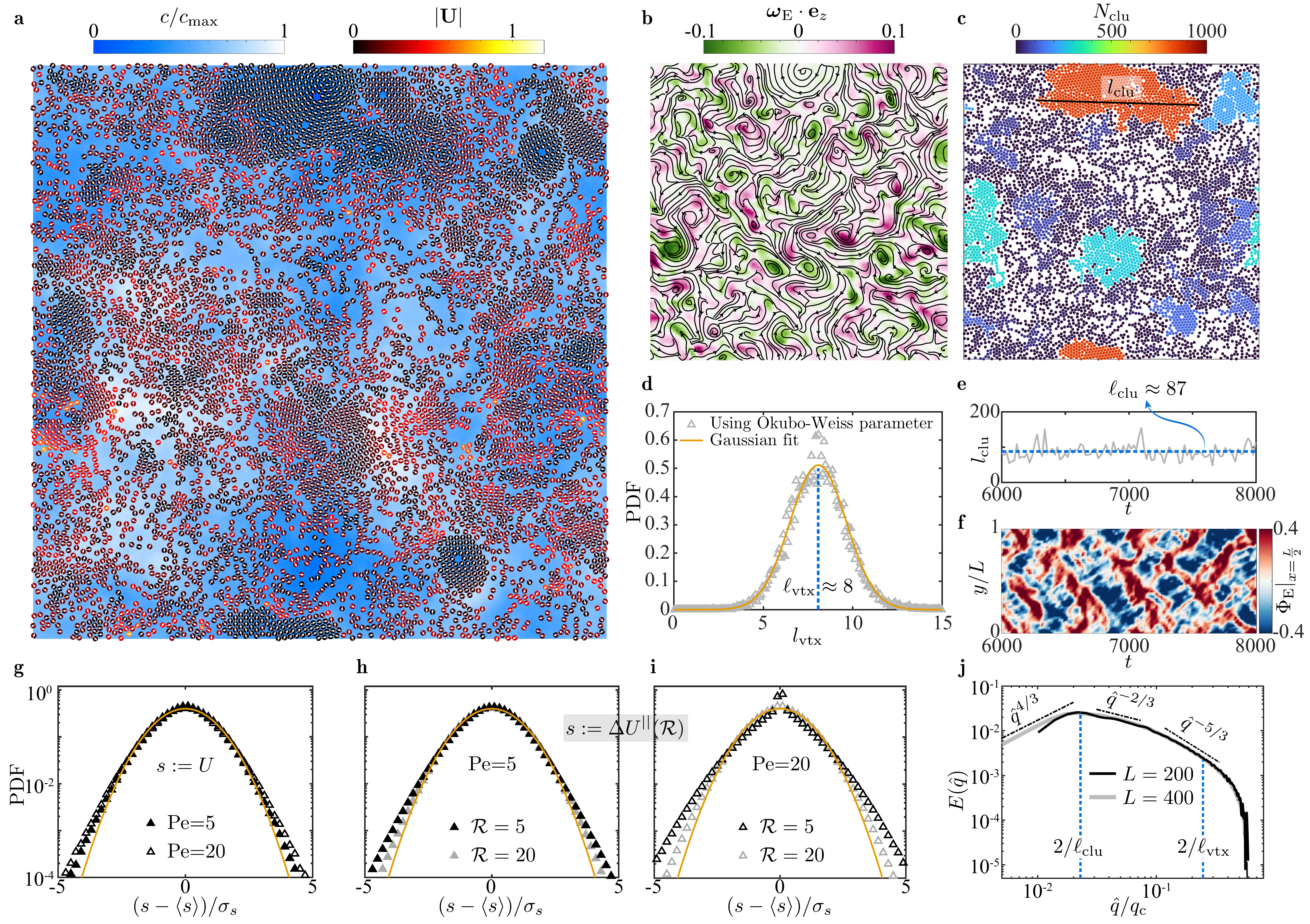}
\caption{
\textbf{Active turbulence.} 
\textbf{a}, instantaneous motion of disks colored by their speed $|\bU|$ and concentration field
$c/\cmax$. Here, $\Pe=20$, $\phi=0.5$, and the domain size is $L=200$. \textbf{b}, vortex-showing streamlines and vorticity component $\bomegaloc \cdot \be_z$ of the continuum flow field $\bUloc$. \textbf{c}, clusters characterized by the number $\Nswarm$ of their constituting disks and the size $\lswarm$ of the largest cluster. \textbf{b} and \textbf{c} showing the full domain are scaled versus \textbf{a}. \textbf{d}, probability density function (PDF) of the vortex size $\lvortex$ with a mean of $\elvortex\approx 8$. \textbf{e}, history of $\lswarm$ and its time-averaged value $\elswarm\approx 87$.
\textbf{f}, kymograph of $\phid|_{x=\frac{L}{2}}$ implies chaos and fading oscillatory patterns. \textbf{g}, PDF of the disk' velocity component $U$. \textbf{h}, PDF of the longitudinal velocity difference $\dUl (\Rsep)$ between two disks separated by a distance $\Rsep$ when $\Pe=5$. \textbf{i}, similar to \textbf{h}, but for $\Pe=20$. In \textbf{g}-\textbf{i}, $\la s \ra$ and $\sigma_s$ denote the average and standard deviation of a random variable $s$, respectively; the curve represents the unit-variance Gaussian function $1/\sqrt{2\pi}\exp(-s^2/2)$. \textbf{j}, kinetic energy spectrum $\Eqnew$ versus the modified wavenumber $\qnew$ (\SI), with an additional dataset incorporated for an expanded domain where $L=400$. Here, $\qc=\pi$ is the characteristic wavenumber corresponding to the disk diameter. 
The left vertical line indicates the size,  $\elswarm \approx 87$ of the largest cluster, while the right line marks the mean vortex size, $\elvortex\approx 8$; both pertain to the case of $L=200$. 
}
\label{fig:5}
\end{figure*}

To confirm this scenario quantitatively, we examine, close to the phase transition, the correlation functions $\gtcfR$ and  $\gocfR$ of the translational and orientational order (\SI).
As illustrated in Fig.~\ref{fig:3}{\textbf{c}}, the former $\gtcfR \propto \Rsep^{-\etat}$ with $\etat\approx 1/25$
when $\Pe=2.3$ (\SI). This algebraic scaling, suggestive of the quasi long-range translational order, typifies a 2D solid, with the power-law exponent $-\etat \geq -1/3$~\cite{mermin1968crystalline}. Conversely, $\gtcfR$  decays exponentially at $\Pe= 2.35$, visually faster than the limiting behaviour $\gtcf \propto \Rsep^{-1/3}$ of solids~\cite{mermin1968crystalline}. The corresponding short-range translational order implies a liquid/hexatic phase when $\Pe \geq 2.35$. Akin to $\gtcfR$, we depict in Fig.~\ref{fig:3}{\textbf{d}} the orientational order correlation function $\gocfR$ that features three scaling laws: at the lowest temperature $\Pe=2.3$, the peaks of $\gocfR$ remain approximately constant, evidencing the long-range orientational order of 2D solids; in the other limit $\Pe=2.5$, $\gocf \propto \exp \lp -\Rsep/40 \rp$ indicates the short-range orientational order possessed by liquids; at intermediate temperatures, $\gocf \propto \Rsep^{-\etao}$, where $-\etao > - 1/4$ at $\Pe=2.35$ and $2.4$ but $-\etao \lessapprox -1/4$ when $\Pe=2.45$. Recalling that the orientational order of a hexatic phase decays algebraically with the exponent $-\etao \geq -1/4$ according to the KTHNY prediction~\cite{halperin1978theory,nelson1979dislocation}, we confirm the presence of a hexatic phase at $\Pe \in [2.35, 2.4]$. Although this specific regime may vary marginally upon doubling the domain size $L$, the overarching two-step melting process remains consistent (\SI).

\subsection*{Instability, Transition, and Active Turbulence}\label{sec:turb}
Viewing the melted disk suspension of $\phi=0.5$ as a continuum active fluid, we examine the spatial-temporal evolution of its Eulerian velocity $\bUloc=\lp \Uloc,\Vloc \rp $ and area fraction $\philoc$ (\SI). Here, $\phid = \lp \philoc-\phi\rp /\phi$ effectively measures the ``pressure'' of the active fluid.

At $\Pe=3$, the domain-averaged ($\la \ra_{xy}$) speed
$\la |\bUloc| \ra_{xy}$ and absolute pressure $\la |\phid| \ra_{xy} $ approach nearly zero after a transient decline (Fig.~\ref{fig:4}{\textbf{b}}), reaching a quiescent flow state (laminar). Conversely, such Eulerian quantities at $\Pe=5$ grow and level off at considerable values, indicating a self-organized active flow of disks (Fig.~\ref{fig:4}{\textbf{a}} and \movRef{6}). The flow appears as large-scale   waves indicated by the time-periodic evolution of the $x$-averaged pressure  $\la \phid \ra_x$ and velocity $\la V \ra_x$ (Fig.~\ref{fig:4}{\textbf{c}}), and the undulating kymograph of $ \phid|_{x=\frac{L}{2}} $ at the middle line $x=\frac{L}{2}$ (Fig.~\ref{fig:4}{\textbf{d}} top). The time evolution of wave (Fig.~\ref{fig:4}{\textbf{e}} top) shares a qualitative similarity with the occurrence of an oscillatory hydrodynamic instability via a Hopf bifurcation---an infinitesimal disturbance growing exponentially saturates to a periodic state. The similarity hints at the resemblance between this activity-induced quiescence-oscillation transition and inertia-triggered  hydrodynamic instabilities, as recognized for polar active fluids~\cite{alert2022active}; that instability occurs when Reynolds ($\Re$) number inversely related to the kinematic viscosity grows. Analogously, increasing $\Pe$ here enables the transition from a more viscous to a less viscous phase (solid-to-liquid-to-gas) as shown in Fig.~\ref{fig:2}, indeed enlarging the effective $\Re$ of the active fluid and thus promoting its instability.

As $\Pe$ is raised from $5$ to $10$, the undulatory dynamics becomes irregular (Fig.~\ref{fig:4}{\textbf{e}} bottom). In parallel,
the kymograph of $ \phid|_{x=\frac{L}{2}} $ exhibits a disrupted wave pattern (Fig.~\ref{fig:4}{\textbf{d}} bottom), seemingly indicative of a secondary instability that is known to trigger the breakdown of streaks and waves in classical hydrodynamics~\cite{schmid2002stability}. Moreover,
the wave breakdown here is reflected by the emergent clusters of disks showcased in Fig.~\ref{fig:4}{\textbf{f}} (\movRef{7}), which imply fluctuations of the effective pressure $\phid$.  The cluster size distribution $\Pswm $ shown in Fig.~\ref{fig:4}{\textbf{g}} further evidences the clustering events at $\Pe=10$ quantitatively (\SI).

At a even higher activity $\Pe=20$, oscillatory patterns fade significantly in favour of emerging vortical structures  (Fig.~\ref{fig:5}{\textbf{a}} and \movRef{8}), hinting the occurrence of active turbulence. Concurrently, enhanced clustering is evident, potentially resulting from the turbulence, as identified in polar active fluids~\cite{worlitzer2021motility}.

Despite their Lagrangian appearance (\SI), Fig.~\ref{fig:5}{\textbf{b}} manifests the vortical structures by the Eulerian flow streamlines and vorticity field $\bomegaloc \cdot \be_z$. This Eulerian description enables using the Okubo-Weiss parameter  (\SI) to quantify the vortex sizes, which are shown in Fig.~\ref{fig:5}{\textbf{d}} to feature a Gaussian probability density function (PDF). Besides, we pinpoint the largest cluster of disks and identify its effective size $\lswarm(t)$ as the longest disk-disk distance (Fig.~\ref{fig:5}{\textbf{c}}). The evolution of $\lswarm$ depicted in Fig.~\ref{fig:5}{\textbf{e}} reveals its time-averaged value $\elswarm \approx 87$.

Apart from the graphic depiction, we compare the statistics of the turbulent disk motion at $\Pe=20$ to those at $\Pe=5$. 
As shown in Fig.~\ref{fig:5}{\textbf{g}}, the PDF of the velocity component $U$ is approximately Gaussian in both cases. Compared to the less active scenario ($\Pe=5$), PDF at $\Pe=20$ departs slightly more from the Gaussian distribution attributed to the corresponding packed and dissipative clusters~\cite{urzay2017multi}. 
Akin to understanding inertial turbulence, 
we examine the velocity difference $\dbU \lp t, \bR, \bRsep \rp = \bU \lp t, \bR +\bRsep \rp - \bU \lp t, \bR \rp  $ between two disks separated by a displacement $\bRsep$, focusing on statistics of its longitudinal component $\dUl = \dbU \cdot \bRsep / |\bRsep|$. Reducing the separation $\Rsep$ from $20$ to $5$, Fig.~\ref{fig:5}{\textbf{h}} \& I suggest that the PDF of $\dUl$ deviates more from the Gaussian profile. The deviation is more pronounced at  $\Pe=20$, as reflected by the fat-tailed distribution for $\Rsep=5$ (Fig.~\ref{fig:5}{\textbf{i}}). The fat tails manifest a stronger probability of high-amplitude extreme events associated with small-scale intermittent processes as well recognized in canonical turbulence~\cite{she1990intermittent,frisch1995turbulence} and also highlighted very recently in a dry active system~\cite{mukherjee2022intermittency}. Here, they are linked to the formation of intense vortical structures yielding a larger relative velocity of two disks.

Focusing on the active turbulence at $\Pe=20$, we display in Fig.~\ref{fig:5}{\textbf{j}} the spectrum $\Eqnew$ of its kinetic energy $\la \bU^2 \ra/2$, versus the modified wavenumber (\SI), $\qnew=2\pi/\lp \Rsep - 2 \rp$. The spectrum scales as $\Eqnew \sim \qnew^{5/3}$ and $\Eqnew \sim \qnew^{-2/3}$, respectively, at small (large $\qnew$) and intermediate length scales. 
Furthermore, we observe a regime with a positive scaling exponent at large length scales (small $\qnew$). Specifically, the scaling $\Eqnew \sim \qnew^{4/3}$ is pronounced for an enlarged domain size of $L=400$. This regime connects with the $\Eqnew \sim \qnew^{-2/3}$ regime via a peak that signifies the maximal energy injection~\cite{giomi2015geometry,alert2020universal}.
Notably, the whole spectrum spanning these three regimes mirrors, in its shape, that of the experimentally observed active nematic turbulence~\cite{martinez2021scaling}. 
In the latter case, the $\Eq \sim q^{-1}$ regime is confined to an intermediate length scale, above a vortex size $\elvortex$, and below a viscous length scale where $\Eq$ peaked. Analogously, our intermediate $\Eqnew \sim \qnew^{-2/3}$ regime is bounded likewise: the lower bound  exceeds the mean vortex size $\elvortex \approx 8$ identified in Fig.~\ref{fig:5}{\textbf{d}}; the upper bound, where the energy spectrum also approximately peaks, aligns closely to the time-averaged size $\elswarm \approx 87$ of the largest cluster (Fig.~\ref{fig:5}{\textbf{e}}).

We note that our simulations have incorporated a weak yet finite fluid inertia of $\Re=0.5$ to approximate Stokes flow (\mat). However, inertia is not the primary factor driving our active turbulence, contrary to other configurations~\cite{goto2015purely,kokot2017active, reeves2021emergence} featuring a dominant inertial effect. Specifically, Ref.~\cite{goto2015purely} demonstrates that a suspension of rotating disks exhibits consistent collective behaviors at varying $\Re$ up to $\approx 0.6$, transitioning to chaos at $\Re \gtrapprox 5$ (see their Figure 7\textbf{c}).

On the other hand, Ref.~\cite{reeves2021emergence} highlights the high sensitivity of the emerging active turbulence to $\Re$ even below $0.1$, unlike the weak $\Re$-dependence depicted here (\SI). To rationalize the discrepancy, we first emphasize that
both settings involve a driving nonlinearity and an auxiliary counterpart. In that study,
an electric field of magnitude $\tE$ above a threshold $\tE_{\text{c}}$ drives an electro-hydrodynamic instability quantified by $\gamma=\tE/\tE_{\text{c}}$. In our setting, the driving nonlinearity arising from the phoretic transport causes a hydrochemical instability characterized by $\Pe$. Despite their different driving mechanisms, both studies feature the same auxiliary nonlinearity: inertia.
The reason why the strong $\Re$-dependency in Ref.~\cite{reeves2021emergence} is absent here stems from a distinction in the strength of the driving nonlinearity relative to the inertial one. 
When the driving nonlinearity is comparable or even weaker than  its inertial counterpart, the driver could feasibly intensify the impact of changing $\Re$, whereas a dominant driving nonlinearity may mitigate this impact.
Ref.~\cite{reeves2021emergence} adopts a nonlinearity level of $\gamma=1.1$, just above the instability threshold: $\gamma=1$, representing a weak driving nonlinearity. 
Conversely, we demonstrate active turbulence at $\Pe=20$, a value far above the threshold, $\Pe \approx 0.5$. Hence, our driving nonlinearity substantially exceeds the inertial counterpart, implying that variations in $\Re$ around unity may wield minimal influence, as indeed supported by additional simulations (\SI).

\section*{Discussion}

By regarding its activity $\Pe$ as the analogue of temperature, we showcase a wet active matter preserving all thermodynamic phases, as previously depicted for dry active matter~\cite{klamser2018thermodynamic,digregorio2018full}. Moving ahead by increasing this activity analogous to Reynolds number, the fluid phase exhibits progressively, a quiescent laminar state, waves via an oscillatory instability, clusters that break the waves down signifying transition, and finally, vortical structures suggestive of active turbulence. This progression highlights a stronger phenomenological resemblance between active and classical fluids in their laminar-turbulent transition than previously recognized.

Controlling both the phase transition and laminar-turbulent transition by activity solely is remarkable, yet not an isolated observation. 
Prior experiments on camphor swimmers have independently demonstrated their crystalline solid state~\cite{soh2008dynamic} and turbulent-like collective activity~\cite{bourgoin2020kolmogorovian},
indeed indirectly supporting our 
our unified interpretation of these dual phenomena.
Their difference stems from the specific range of phoretic activity: lower activity prompts crystallization, while higher activity induces turbulence (\SI), corroborating our predictions (Fig.~\ref{fig:2}). Ultimately, this discovery offers a paradigm to optimize the reconfigurability and functionality of active systems using minimal control.

The observed capability to discern both experimentally identified transitions is attributed to the simultaneous integration of long-range hydrodynamic and chemical interactions.  Studies focusing solely on chemical interaction reproduced hexagonal crystallization but failed to capture active turbulence~\cite{gouiller2021two}. Conversely, simulations considering only hydrodynamic interaction illustrated turbulence but missed crystallization~\cite{zantop2022emergent,qi2022emergence}. This comparative analysis highlights the unique predictive power unlocked by concurrently addressing both hydrodynamic and chemical interactions in modelling chemically active fluids---an aspect that merits broader recognition within the community.

\newpage
\section*{Methods}
A more detailed description of the materials and methods is provided in \SI.

\subsection*{Mathematical Model and Governing Equations}
We adopt the minimal physicochemical hydrodynamic model of IPA~\cite{michelin2013spontaneous}. 
Without considering its internal flow, the IPA has been modeled as a three-dimensional spherical particle or a 2D circular disk~\cite{hu2019chaotic}. The Marangoni interfacial flow is represented by a solute-induced tangential velocity at the surface of particles or disks. The simple concept makes it a popular reference model for researching the behavior of a single IPA~\cite{hu2019chaotic,chen2021instabilities,kailasham2022dynamics,zhu2023self} or a pair~\cite{nasouri2020exact,chen2024buoyancy}. Most importantly, it retains the critical signature of an IPA, \viz exploiting the convection of a chemical product to sustain a Marangoni propelling flow upon spontaneous symmetry-breaking. Concomitantly, the signature scenarios of IPAs observed in experiments~\cite{hokmabad2021emergence,dwivedi2021solute} can be captured by the model~\cite{hu2019chaotic,lin2020direct,chen2021instabilities}: by increasing $\Pe$, it transitions from a stationary state to steady propulsion, and further to chaotic motion. 

We describe the hydrochemical model by the dimensional Stokes equation for the velocity $\tbu$ and pressure $\tp$, along with the advection-diffusion counterpart for the solute concentration $\tc$ of a chemical species:
\begin{subequations}
\begin{align}
    \tgrad \cdot \tbu =0, \quad    \tgrad \tp = \mu \tgrad^2 \tbu,\\
    \frac{\partial \tc}{\partial \tt} + \tbu \cdot \tgrad \tc  = \mD \tgrad^2 \tc,
\end{align}
\end{subequations}
where $\mu$ is the dynamic viscosity of the solvent. 

In addition to the fluid motion and solute transport, the model encodes the physicochemical activity at the disk surface by specifying its boundary conditions. This involves two processes: first, the uniform and constant solute emission from its surface  
\begin{align}\label{eq:bc_c_dim}
    \mD \bn \cdot \tgrad \tc & = -\mA;
\end{align}
second, the generation of a slip velocity by a local solute gradient
\begin{align}\label{eq:bc_up_dim}
    \tbu_{\text{slip}} & = \mM \lp \bI -\bn\bn \rp \cdot \tgrad \tc
\end{align}
tangential to the surface. 

The dimensionless equations are 
\begin{subequations}\label{eq:main}
\begin{align}
    \grad \cdot \bu = 0, \quad \grad p = \grad^2 \bu,\label{eq:stokes}\\
    \frac{\partial c}{\partial t} + \bu \cdot \grad c  =\frac{1}{\Pe} \grad^2 c.
\end{align}
\end{subequations}
Here, we have chosen $a$, $\mA \mM /\mD$, $\mu \mA \mM /\lp \mD a\rp$, and $\mA a/\mD$ as the characteristic length, velocity, pressure, and concentration scales, respectively, \eg $\bu = \tbu/\lp \mA \mM /\mD \rp $ and $c=\tc/\lp \mA a/\mD\rp$. 
The dimensionless versions of \eqref{eq:bc_c_dim} and \eqref{eq:bc_up_dim}
read
\begin{subequations}\label{eq:bc}
\begin{align}
    \bn \cdot \grad c & = -1,\\
    \bu_{\text{slip}} & = \lp \bI -\bn\bn \rp \cdot \grad c. \label{eq:bc_u}
\end{align}
\end{subequations}
The slip velocity Eq.~\eqref{eq:bc_u} allows the disk to freely move with a translational velocity $\bU$ and a rotational velocity $\bOmega=\Omega \be_z$ with $\be_z = \be_x \times \be_y$. Note that \eqref{eq:bc_u} is not the boundary condition for $\bu$, which involves $\bU$ and $\Omega$ as detailed in the SI.

It is worth-noting that our numerical implementation does not exactly solve the Stokes equation \eqref{eq:stokes} but rather the Navier-Stokes equation featuring a finite yet small Reynolds number $\Re=\mA\mM a/\lp \nu \mD\rp$, where $\nu$ is the kinematic viscosity of the solvent. We choose $\Re=0.5$ throughout this study. The associated inertia term is needed for time-marching the momentum equation. We find that its influence on the dynamics of a phoretic disk is reasonably weak, as discussed in the numerical validation section of \SI.
In addition, we have also shown in the \SI~    that variations in inertia within the range $\Re\in [0.1, 2]$ do not significantly alter the critical features of the collective phenomenon, including vortex size, velocity statistics, and the energy spectrum.

Unless otherwise specified, the size of the doubly-periodic domain is $L=100$. In certain configurations, $L=200$ and $L=400$ are adopted.

\subsection*{Numerical Simulations}\label{sec:numerical}
We numerically solve the dimensionless equations \eqref{eq:main} in a periodic square domain of size $L$, subject to the boundary conditions \eqref{eq:bc} at the surfaces of freely moving disks. We have adapted a massively parallel flow solver~\cite{jiang2019boundary,jiang2022simple} to cater for our physicochemical hydrodynamic configuration. The solver employs a lattice Boltzmann method to resolve fluid motion and solute transport while integrating an immersed boundary method to represent the surfaces of finite-sized disks. The implementation and validations are presented in \SI. 

Using four computer nodes with four Intel Xeon Gold 6230R CPUs and 104 cores each, a typical simulation runs for several days to weeks, depending on $\Pe$ and $L$. For example, active turbulence tends to emerge at a high $\phi$ values, corresponding to a larger number of  disks. Moreover, to broaden the energy spectrum of active turbulence, we use a domain of size $L=200$ that doubles the default size, and even $L=400$ for a more definitive insight (see Fig.~\ref{fig:5}{\textbf{j}}). These cases typically run for weeks using nine nodes to attain statistically invariant transitional or turbulent dynamics.

We have studied a few phoretic swimmers, as elaborated in the SI and illustrated through  Supplementary Videos 9-11. Numerical simulations have successfully captured the key experimental observations on a pair of interacting active droplets~\cite{hokmabad2021emergence}.

\section*{Acknowledgments}
We thank the helpful discussions with Ricard Alert, Cecile Cottin-Bizonne, Marjolein Dijkstra, Amin Doostmohammadi, Gerhard Gompper, Endao Han, Hisay Lama, Gaojin Li, Detlef Lohse, S\'ebastien Michelin, Alexander Morozov, Ran Ni, Ignacio Pagonabarraga, Fernando Peruani, Kai Qi, Francesc Sagu\'es, Chunlei Song, Christophe Ybert, and Guangpu Zhu, as well as Wei-Fan Hu for sharing with us their data used in the validation. The authors thank the anonymous reviewers for their valuable insights and constructive comments, especially those regarding the kinetic energy spectrum.
Q.Y. is supported by the research scholarship from the National University of Singapore and the China Scholarship Council. L.Z. acknowledges the Singapore Ministry of Education Academic Research Fund Tier 1 grant (A-8000197-01-00). Some computation of the work was performed on resources of the National Supercomputing Centre, Singapore (https://www.nscc.sg), as well as the supercomputer Fugaku provided by RIKEN through the HPCI System Research Project (Project ID: hp230185).

\clearpage
\renewcommand\appendixpagename{Supplementary Information}
\appendix
\appendixpage

\renewcommand{\appendixname}{\SI}
\input{SI_main_post_accept1}

\clearpage

\end{document}

%% file: sym_post_accept1.tex
\def \TITLEnew {Shaping active matter from crystalline solids to active turbulence}
\def \SI {SI}
\def \viz {viz.,~}
\def \ie {i.e.,~}
\def \eg {e.g.,~}
\newcommand{\figref}[2][{}]{Fig.\
\ref{#2}\ifthenelse{\isempty{#1}}{}{\,(#1)}}
\newcommand{\figrefS}[2][{}]{Fig.~S\ref{#2}\ifthenelse{\isempty{#1}}{}{\,(#1)}} 
\newcommand{\eqnref}[1]{Eq.~\eqref{#1}}
\newcommand{\eqnrefS}[1]{Eq.~S\eqref{#1}}
\newcommand{\movref}[1]{Movie S{#1}}
\newcommand{\movRef}[1]{Supplementary Video {#1}}
\def \mat {Methods}
\def \jnc {Nature Communications}
\def \edtnc {Editors of \jnc}
\def \editor {\edtnc}
\def \journal {\jnc}

%% file: sym_math_post_accept1.tex
\def \Pe {\text{Pe}}
\def \Re {\text{Re}}
\def \Ma {\text{Ma}}

\def \dirac {\delta}
\def \diracD {\dirac_{\text{dis}}}

\def \Pec {\Pe_{\text{col}}}
\def \Pecri {\Pe^{\ast}}

\def \tgrad {\tilde{\grad}}

\def \mA {\mathcal{A}}
\def \mD {\mathcal{D}}
\def \mC {\mathcal{C}}
\def \mE {\mathcal{E}}
\def \mP {\mathcal{P}}
\def \mV {\mathcal{V}}

\def \mEd {\mE_{\text{d}}}
\def \mEf {\mE_{\text{f}}}
\def \mDf {\mD_{\text{f}}}
\def \mVf {\mV_{\text{f}}}

\def \mPslip {\mP_{\text{slip}}}
\def \mProt {\mP_{\text{rot}}}

\def \ba {\mathbf{a}}
\def \bb {\mathbf{b}}
\def \bc {\mathbf{c}}
\def \bd {\mathbf{d}}
\def \bg {\mathbf{g}}
\def \bh {\mathbf{h}}
\def \bi {\mathbf{i}}

\def \be {\mathbf{e}}
\def \bq {\mathbf{q}}

\def \bu {\mathbf{u}}
\def \br {\mathbf{r}}
\def \hbr {\hat{\br}}
\def \hg {\hat{g}}

\def \bsigma {\boldsymbol{\sigma}}
\def \bR {\mathbf{R}}
\def \bU {\mathbf{U}}
\def \tbU {\tilde{\bU}}
\def \bF {\mathbf{F}}
\def \bmF {\boldsymbol{\mathcal{F}}}
\def \bf {\mathbf{f}}
\def \bL {\mathbf{L}}
\def \bat {\mathbf{a}_{\mathrm{t}}}
\def \bar {\mathbf{a}_{\mathrm{r}}}
\def \bOmega {\boldsymbol{\Omega}}
\def \hbe {\hat{\be}}
\def \htau {\hat{\tau}}
\def \feq {f^{\text{eq}}}

\def \rhob {\rho_0}
\def \bn {\mathbf{n}}
\def \bI {\mathbf{I}}
\def \bRL {\bR_{\text{L}}}

\def \tbu {\tilde{\bu}}
\def \tbr {\tilde{\br}}
\def \tbR {\tilde{\bR}}

\def  \C {\text{C}}
\def  \d {\text{d}}
\def \i {\text{i}}

\def \tt {\tilde{t}}
\def \tp {\tilde{p}}
\def \tc {\tilde{c}}
\def \tx {\tilde{x}}
\def \ty {\tilde{y}}
\def \tE {\tilde{E}}

\def \buslip {\bu_{\text{slip}}}
\def \tbuslip {\tbu_{\text{slip}}}
\def \bup {\bu_{\text{p}}}

\def \mM {\mathcal{M}}
\def \mR {\mathcal{R}}
\def \mI {\mathcal{I}}

\def \lp {\left(}
\def \rp {\right)}

\def \aUrms {U_{\text{rms}}}
\def \avel {\aUrms}
\def \avelt {\overline{\aUrms}}

\def \kp {k^{\prime}}

\def \tell {\tilde{\ell}}

\def \bre {\br^{\text{E}}}
\def \brl {\br^{\text{L}}}
\def \cp {c^{\mathrm{int}}}
\def \cc {c^{\text{c}}}
\def \mp {m^{\mathrm{int}}}
\def \mr {m^{\mathrm{pre}}}

\def \uint {\bu^{\mathrm{int}}}

\def \dr {\d r}

\def \xc {x_{\mathrm{c}}}
\def \yc {y_{\mathrm{c}}}
\def \Xc {X_{\mathrm{c}}}
\def \Yc {Y_{\mathrm{c}}}

\def \as {\tilde{a}_{\mathrm{s}}}

\def \tU {\tilde{U}}
\def \ttheta {\tilde{\theta}}
\def \tbe {\tilde{\mathbf{e}}}
\def \tB {\tilde{B}_1}

\def \Rc {\mathcal{R}_{\text{c}}}
\def \nrad {n_{\text{rad}}}

\def \Rgap {\mathcal{R}_{\mathrm{gap}}}
\def \epar {\mathbf{e}_{\mathcal{Y}}}
\def \epen {\mathbf{e}_{\mathcal{X}}}
\def \Upar {V}
\def \Upen {U}

\def \qc {q_{\mathrm{c}}}

\def \mbR {\boldsymbol{\mathcal{R}}}
\def \mR {\mathcal{R}}
\def \mV {\mathcal{V}}

\def \Rsep {\mR}
\def \bRsep {\mbR}
\def \hRsep {\hat{\Rsep}}

\def \Rsepprime {\Rsep^{\prime}}
\def \Rseplb {\Rsep_{\rm lb}}
\def \Rsepsft {\Rsep_{\rm sft}}
\def \Rsepnew {\hRsep}

\def \qnew {\hat{q}}

\def \dbU {\Delta \bU}
\def \dU {\Delta U}
\def \dUl {\dU^{\parallel}}
\def \dUt {\dU^{\perp}}

\def \MSD {\mathrm{MSD}}
\def \PDF {\mathrm{PDF}}
\def \bRd {\mathbf{R}^{\mathrm{d}}}

\def \cmax {c_{\text{max}}}

\def \elvis {\ell_{\mathrm{vis}}}

\def \elvortex {\ell_{\mathrm{vtx}}}
\def \elswarm {\ell_{\mathrm{clu}}}

\def \lvortex {l_{\mathrm{vtx}}}
\def \lswarm {l_{\mathrm{clu}}}

\def \avelswarm {\overline{\ell}_{\mathrm{clu}}}

\def \Eq {E(q)}
\def \Eqnew {E(\qnew)}

\def \Nnb {\mathcal{N}}

\def \tran {\mathbf{q}_{0}}
\def \gtcf {g_{\tran}}
\def \gocf {g_6}
\def \gtcfR {\gtcf\lp \Rsep \rp}
\def \gocfR {\gocf\lp \Rsep \rp}

\def \etat {\eta}
\def \etao {\eta^{\prime}}

\def \gUUt {g_{UU}(\tau)}

\def \cloc {c_{\loc}}

\def \Nswarm {N_{\text{clu}}}

\def \thetas {\vartheta}
\def \thetac {\varphi}
\def \la {\langle}
\def \ra {\rangle}

\def \loc {\mathrm{E}}

\def \Lloc {L_{\loc}}
\def \philoc {\phi_{\loc}}
\def \phid {\Phi_{\loc}}

\def \bUloc {\mathbf{U}_{\loc}}
\def \Uloc {U_{\loc}}
\def \Vloc {V_{\loc}}
\def \Nloc {N_{\loc}}
\def \mVloc {\mV_{\loc}}
\def \bomegaloc {\boldsymbol{\omega}_{\loc}}

\def \Sins {\mathcal{S}_k}
\def \gwwR {g_{\omega \omega}(\Rsep)}
\def \FFT {\mathcal{F}}
\def \iFFT {\mathcal{F}^{-1}}
\def \OW {\mathrm{OW}}
\def \CIR {\Gamma}
\def \parx {\partial_x}
\def \pary {\partial_y}
\def \becir {\mathbf{e}_{\mathrm{cir}}}
\def \Rcir {\Rsep_{\mathrm{cir}}}
\def \Svtx {\mathcal{S}_{\mathrm{vtx}}}
\def \Ustp {U_{\mathrm{strip}}}
\def \Vstp {V_{\mathrm{strip}}}

\def \Pswm {\mathcal{P}(\Nswarm)}
\def \Nswmcut {\Nswarm^{\nmid}}

\def \Ca {\mathcal{C}_1}
\def \Cb {\mathcal{C}_2}

\def \rhop {\rho_{\mathrm{o}}}
\def \rhof {\rho}
\def \muf {\mu}
\def \Fp {\tilde{\mathbf{F}}}
\def \Lp {\tilde{\mathbf{L}}}
\def \mmp {\tilde{m}}
\def \bup {\bu^{\text{int}}}
\def \Vp {\tilde{V}}
\def \Ip {\tilde{I}}
\def \ap {\frac{\d \Up}{\d \tilde{t}}}
\def \op {\frac{\d \Op}{\d \tilde{t}}}
\def \Up {\tilde{\mathbf{U}}}
\def \Op {\tilde{\boldsymbol{\Omega}}}

\def \Uc {\tilde{U}}

\def \apscl {\frac{\d \bU}{\d t}}
\def \opscl {\frac{\d \bOmega}{\d t}}

  \def \dtp {\hat{\delta}}

\def \transp {\mathsf{T}}

\def \gUUR {g_{UU}(\Rsep)}

%% file: SI_main_post_accept1.tex
\captionsetup{labelformat=custom}

In this document, we provide supplementary information (SI) to complement the main article. Initially, we clarify the definitions of specific variables in Sec.~\ref{sec:def}. Following this, the numerical implementation and validation processes are detailed in Sec.~\ref{sec:numerical_SI}. Sec.~\ref{sec:theory} elaborates on the theoretical prediction of the phase transition.

Section~\ref{sec:interaction} explores typical scenarios involving the interaction of two phoretic disk swimmers and characterizes their hydrodynamic interactions. The extension from two disks to multiple disks forming chains is discussed in Sec.~\ref{sec:chain}.  Additional data pertaining to two-dimensional melting are presented in Sec.~\ref{sec:melting}. Subsequently, Sec.~\ref{sec:turbulence} delves into certain subtle aspects of active turbulence.

In Sec.~\ref{sec:clustering}, we delineate the distinctions between the microswimmer clusters observed in our study and those documented in prior research. Sec.~\ref{sec:reconciling} is dedicated to elucidating how our findings bridge the gap between two contrasting experimental observations on camphor surfers. Lastly, the behavior of isotropic phoretic agents (IPAs) in an unbounded domain is discussed.

Variables here are dimensionless unless otherwise mentioned. Dimensional variables indicated by ~$\tilde{}$~ as in the main article appear in Sec.~\ref{sec:theory} only.

\section{Supplementary Notes}\label{sec:def}
Here, we present the variables not elaborated on in the main article, organizing them based on their initial appearance in the corresponding figures. 
\subsection{Fig. 1: Numerical Setup and Diverse Collective Behaviors}
The root-mean-square velocity $\aUrms(t)$ of  $N$ disks is:
\begin{align}
    \aUrms(t) =\sqrt{\frac{1}{N}\sum^N_{k=1} |\bU_k(t)|^2},
\end{align}
where $\bU_k(t)$ denotes the translational velocity of the $k$-th disk at time $t$. 

We define the pair correlation function $g(\Rsep)$ as a function of the center-to-center distance $\Rsep$ between a pair of disks:
\begin{align}
g(\Rsep) &= \frac{\sum^N_{k=1} \sum_{\kp \neq k} \diracD \lp \Rsep, \Delta \Rsep, |\bR_{\kp} -\bR_k| \rp }{\pi\Rsep \Delta \Rsep N \lp N-1 \rp/L^2} , 
\end{align}
where $\bR_k$ is the positional vector of  disk $k$ and $\Delta \Rsep$ is a sufficiently small increment of $\Rsep$.
Here, $\diracD (\Rsep,\Delta \Rsep,\Rsep^{\prime}) $ is the discrete  Dirac delta function defined below 
\begin{align}
       \diracD (\Rsep,\Delta \Rsep,\Rsep^{\prime}) = 
\begin{cases}
    1,& \text{if  }\;   \Rsep  \leq \Rsep^{\prime} \leq \Rsep +\Delta \Rsep,\\
    0,              & \text{otherwise}.
\end{cases}
\end{align}

The mean square displacement (MSD) as a function of the time lag $\tau$, can be expressed as:
\begin{align}
    \MSD(\tau) = \left \langle \overline{\overline{|\bRd(t+\tau)-\bRd(t)|^2}}  \right \rangle, 
\end{align}
where $\bRd(t)=\int_0^t \bU(t^{\prime})\mathrm{d}t^{\prime}$ is the total displacement of a disk at time $t$ with respect to its initial position. Here, $\langle\bullet\rangle$ indicates averaging over all disks as will be adopted throughout this work. Also, $\overline{\overline{\lp \bullet \rp }}$ means averaging over a series of moving time windows. We note that $\bRd$ is distinct from $\bR$, where the former has not been shifted to be bounded in the periodic domain, unlike the latter.

\subsection{Fig. 2: Phase Diagram and Theoretical Scaling}
In Fig. 2 of the main article, we show the time-averaged root-mean-square velocity $\avelt \lp \phi, \Pe \rp$. Here, $\overline   {\lp \bullet \rp } $ means time-averaging, a convention consistently used throughout this work.

\subsection{Fig. 3: Defect-mediated Melting }
We introduce the two-dimensional (2D) static structure factor $S(\bq)$ varying with the wave vector $\bq = q\exp\lp \i \theta \rp $
\begin{align}
    S(\bq) &= \frac{1}{N}\overline{I (\bq) I (-\bq)}, 
\end{align}
where
\begin{align}
    I (\bq) &= \sum_{k=1}^{N}\exp(\i \bq \cdot \bR_k), 
\end{align}
and $\i$ represents the imaginary unit.

The local translational  $\psi_{\bq_0}(\bR_k)$ and orientational order $\psi_6(\bR_k)$ are:
\begin{subequations}
\begin{align}
    \psi_{\bq_0}(\bR_k) &= \exp(\i\bq_0\cdot \bR_k), \\
    \psi_6(\bR_k) &= \frac{1}{\Nnb_k}\sum_{j\in \Nnb_k} \exp(6 \i \varphi_{k j}),
\end{align}
\end{subequations}
where $\bq_0$ is the wave vector at the maximum of the first diffraction peak of $S(\bq)$, and $\Nnb_k$ is the number of disk $k$'s neighbours, identified through  Voronoi tessellation. Besides, $\varphi_{kj}$ measures the orientation of the $j$-th neighbour with respect to the host, \viz disk $k$, that is, the angle between their center-to-center displacement vector and a fixed reference orientation.
In this case, we choose the basis vector $\be_x$ as the reference without loss of generality.

The translational or orientational correlation function $g_{\alpha}(\Rsep)$ with $\alpha \equiv \bq_0,6$ is given by:
\begin{align}
     &g_{\alpha}(\Rsep) =  \nonumber \\
     &\frac{\overline{\sum^N_{k=1} \sum_{k^{\prime} \neq k} \psi_{\alpha}^*(\bR_{k^{\prime}}) \psi_{\alpha}(\bR_{k})  \diracD (\Rsep -|\bR_{k^{\prime}} - \bR_k|)}}{\overline{\sum^N_{k=1}\sum_{k^{\prime}\neq k}\diracD(\Rsep -|\bR_{k^{\prime}} - \bR_k|)}}, 
\end{align}
where the asterisk $^{\ast}$ denotes the complex conjugate operator.

\subsection{Fig. 4: Instability and Transition of Active Fluids}

\begin{suppfigure}[tbh!]
    \centering
    \includegraphics[width=1\linewidth]{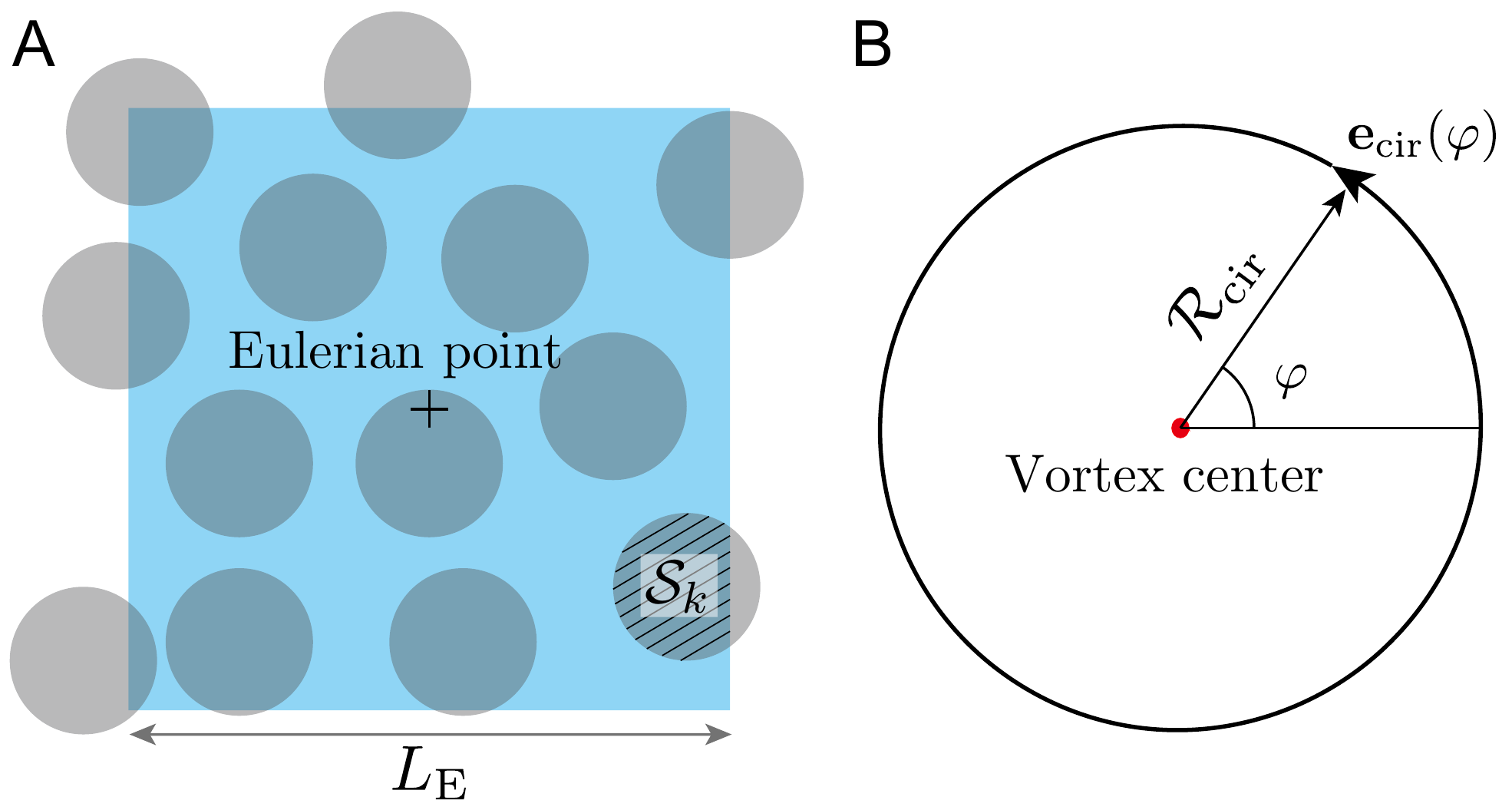}
    \caption{A, calculating   the local variables, \eg area fraction $\philoc$ of disks (circles) at a given Eulerian position ($+$) via spatially sampling a square window. $\Sins$ denotes the area of disk $k$ within the window. B, calculating the scaled circulation on a circle of radius $\Rcir$. The red dot represents a candidate vortex center, $\thetac$ is the polar angle, and $\becir(\thetac)$ is the tangential unit vector.}
    \label{fig:locfield_vortex}
\end{suppfigure}

We consider the disks as atoms or molecules of a new active matter or phase, and map the phase transition onto the classical thermodynamic version. For a fluidized active matter of such, \viz active fluid manifested in Figs. 4 and 5 of the main article, we can study its continuum dynamics via an Eulerian approach in addition to the Lagrangian counterpart, \ie examining the motion of individual disks. 
This resembles adopting the canonical macroscopic description of fluid motion without considering the microscopic molecular motion. In the same vein, we take the macroscopic, Eulerian variable $\mVloc$ as the spatial average of the microscopic, Lagrangian variable $\mV$, such as $\phi$ and $\bU$ of disks.

For a given Eulerian position, we sample a square window of size $\Lloc$ centred at that position (\figrefS{fig:locfield_vortex}{A})  and perform area-weighted averaging of $\mV$ to obtain
\begin{align}
\displaystyle \mVloc = \frac{1}{\Lloc^2}
\begin{cases}
     \sum^{\Nloc}_{k}  \Sins, & \text{if  }\;   \mV:=\phi, \\
    \sum^{\Nloc}_{k} \mV \Sins,              & \text{otherwise}.
\end{cases}
\end{align}
Here, $\Nloc$ is the number of disks that overlap with the window, and $\Sins$ is the overlapping area of disk $k$. To ensure sufficient statistical confidence, we choose a window size $\Lloc = \sqrt{10 \pi /\phi }$ such that $\Nloc \approx 10$ on average. Using the calculated continuum velocity $\bUloc=\Uloc \be_x + \Vloc\be_y$, we can also obtain the corresponding vorticity $\bomegaloc = \grad \times \bUloc$.

Likewise, we calculate the Okubo-Weiss ($\OW$) parameter \cite{okubo1970horizontal, weiss1991dynamics},
\begin{align}
\OW=(\frac{\partial \Uloc}{\partial x} + \frac{\partial \Vloc}{\partial y})^2 - 4 \frac{\partial \Uloc}{\partial x} \frac{\partial \Vloc}{\partial y} + 4 \frac{\partial \Uloc}{\partial y} \frac{\partial \Vloc}{\partial x}
\end{align}
to help identify the vortical structures in our system. 
To do so, we first locate all the Eulerian grid points where $\OW<0$. Next, for each of these found points, we calculate the scaled circulation 
\begin{align}
\CIR = \frac{1}{2\pi} \oint
\frac{\bUloc}{|\bUloc|} \cdot \becir(\thetac) \d\thetac
\end{align}
on a circle of radius $\Rcir$ centered there (\figrefS{fig:locfield_vortex}{B}).  Here, $\thetac$ is the polar angle, and $\becir \lp \thetac \rp = -\cos\thetac\be_x + \sin\thetac \be_y$ is the tangential vector on the circle. We choose $\Rcir=3$, neglecting the vortical structures smaller than that size. This choice seems reasonable given the disk diameter of $2$. Ideally, $|\CIR|=1$ for a grid point that coincides with the center of a perfect circular vortex. In practice, we consider a particular point as a vortex center when $|\CIR| \geq 0.85$. Having probed the center of vortical structures, we further determine their occupied area $\Svtx$. For a probed center of vortex, we examine $\bomegaloc$ and $\OW$ at the Eulerian points within a circular domain of radius $20$. Most of them would share the same sign of $\bomegaloc$, and those with $\OW<0$ are counted for calculating the area $\Svtx$. Accordingly, we obtain the effective radius $\lvortex=\sqrt{\Svtx/\pi}$ of the identified vortex. 
 
The clusters of disks shown in Figs.~4F and 5C of the main article are identified by the following criterion: the surface-to-surface distance between every two neighbouring disks of a cluster is below $0.15$.  We then follow Ref.~\cite{theers2018clustering} to calculate their size distribution function $\Pswm$, 
\begin{align}
    \Pswm = \frac{\Nswarm}{N}\mathcal{G}(\Nswarm), 
\end{align}
where $\Nswarm$ is the number of disks constituting a cluster, and $\mathcal{G}(\Nswarm)$ is the number of clusters sized $\Nswarm$. The distribution has been normalized to fulfil 
\begin{align}
\sum^N_{\Nswarm=1} \Pswm=1.
\end{align}

In Fig. 4\textbf{g} of the main article, we fit the calculated cluster size distribution following Ref.~\cite{bonabeau1999scaling}, 
\begin{align}
\Pswm=\Ca \Nswarm^{-\Cb} \exp \lp -\Nswarm/\Nswmcut\rp,
\end{align}
where the prefactor $\Ca$, exponent $\Cb$, and cutoff size $\Nswmcut$ are fitting parameters. The emergence of clustering can be characterized by $\Nswmcut$, which is approximately $17$ for $\Pe=5$, and $155$ when $\Pe=10$.

\subsection{Fig. 5: Active Turbulence}
\subsubsection{Energy spectrum for point microswimmers}

This study examines active turbulence of finite-sized microswimmers, the calculation of the kinetic energy spectrum differs slightly from that for point microswimmers.

To start with, we describe how the spectrum $\Eq$ versus the wavenumber $q=2\pi/\Rsep$ is calculated for point microswimmers. First, we obtain the equal-time two-point velocity correlation function $g_{UU}(\bRsep)$ following
\begin{align}\label{eq:g_UU}
     & g_{UU}(\bRsep) = \nonumber \\ & \frac{\overline{\sum^N_{k=1} \sum_{k^{\prime} \neq k} \bU_{k^{\prime}}(t) \cdot \bU_k(t)  \diracD[\bRsep -(\bR_{k^{\prime}} - \bR_k)]}}{\overline{\sum^N_{k=1}\sum_{k^{\prime}\neq k}\diracD[\bRsep -(\bR_{k^{\prime}} - \bR_k)]}}. 
\end{align}
Then, applying the Fourier transform of $g_{UU}(\bRsep)$,
\begin{align}\label{eq:Eq}
    E(q)&=\left \langle \frac{q}{2\pi}\int   \exp(-\i\bq\cdot \bRsep)g_{UU}(\bRsep)\mathrm{d}^2\bRsep \right \rangle_{\theta},     
\end{align}
where $\langle \rangle_{\theta}$ denotes averaging over the phase angle $\theta$ of the wave vector $\bq$.

Instead of using Eq.~\eqref{eq:Eq}, there is another alternative method. It lies in an angular average of $g_{UU}(\bRsep)$ resulting in the one-dimensional velocity correlation function $\gUUR = \langle g_{UU}(\bRsep) \rangle_{\theta}$ with $\Rsep = |\bRsep|$. Accordingly, the energy spectrum can be obtained by~\cite{nishiguchi2015mesoscopic}:
\begin{align}\label{eq:PSD1D}
    \Eq =q \int_0^{\infty} \gUUR \Rsep \ J_0(q\Rsep) \d \Rsep,
\end{align}
where $J_0$ is the zeroth-order Bessel function of the first kind. 
Notably, the second method assumes an isotropic active turbulence. 
Our analysis of the data reveals that the two expressions yield nearly the identical energy spectra with indiscernible differences.

\subsubsection{Energy spectrum for finite-sized
microswimmers}
Eqs.~\eqref{eq:Eq} and \eqref{eq:PSD1D} can be directly applied to active turbulence in point microswimmers, where the inter-swimmer distance $\Rsep$ features a lower bound of zero. Similarly, they can also be used to calculate the energy spectrum of inertial turbulence, because the velocity correlation function $\gUUR$ is defined for any two positions in space, regardless of their proximity. Nevertheless, this scenario is altered for finite-sized swimmers, where the minimum inter-swimmer distance $\Rsep$ is not zero but a finite value $\Rseplb$, \viz the lower bound. Consequently,  $\gUUR$ is not defined when $\Rsep \in [0, \Rseplb ) $, which results in the abnormal energy spectrum $\Eq$ calculated from $\gUUR$ via Eq.~\eqref{eq:PSD1D}. This anomaly implies a subtle incompatibility between the classical turbulence theory with the active turbulence in finite-sized active particles.

In fact, previous studies~\cite{qi2022emergence,zantop2022emergent} on active turbulence in finite-sized microswimmer have 
reported the ``artifact'' in $\Eq$ based on Eq.~\eqref{eq:PSD1D}. Both groups overcome this artifact by shifting the correlation function  $g_{UU}(\Rsep)$ along the $\Rsep$ axis, yielding a shifted correlation function $\hg_{UU}(\Rsepnew)$ versus a modified inter-particle distance $\Rsepnew$. Here, $\Rsepnew$ effectively indicates the minimum surface-to-surface distance between two swimmers. 
Upon such a shift, the lower limit of $\Rsepnew = \Rsep - \Rseplb$ becomes zero, which permits using the canonical energy spectrum formula.

Despite circumventing the anomaly or artifact in the energy spectrum, this shifting approach introduces a side effect. The new velocity correlation function $\hg_{UU}(\Rsepnew)$ effectively measures the correlation of two finite-sized swimmers versus their minimum surface-to-surface distance $\Rsepnew$. This definition of distance differs from the center-to-center distance between, either two point swimmers showing active turbulence, or two spatial positions in inertial turbulence. Consequently, the energy spectrum calculated from $\hg_{UU}(\Rsepnew)$ features a new, `shifted' wavenumber 
\begin{align}\label{eq:qnew}
\qnew = \frac{2\pi}{\Rsepnew} = \frac{2\pi}{\Rsep - \Rseplb},     
\end{align}
in contrast to the original one $q = 2\pi / \Rsep$.  In our case, the minimum center-to-center distance between swimmers is $\Rseplb = 2$, \ie two radii of phoretic disks, leading to $\qnew = 2\pi/\lp \Rsep-2 \rp$. 

In this work, we have adopted the shifting approach and shown the energy spectrum $\Eqnew$ versus the shifted wavenumber $\qnew$ 
scaled by a characteristic wavenumber $\qc$. Here,  $\qc=2\pi/2=\pi$ is defined based on the diameter $2$ of disks.

\section{Supplementary Methods}\label{sec:numerical_SI}
\subsection{Governing Equations}\label{sec:equation_SI}
To address our hydrochemical problem, we adapt a massively parallel flow solver using a Lattice Boltzmann method (LBM)~\cite{chen1998lattice} and an immersed boundary method (IBM)~\cite{peskin1972flow, mittal2005immersed}. It is worth-noting that LBM solving hydrochemical dynamics was recently used to study suspensions of Janus colloids in a Hele-Shaw cell~\cite{scagliarini2020unravelling}. Besides LBM, the multi-particle collision dynamics method~\cite{gompper2009multi} has also been employed to investigate hydrochemical interactions within Janus collectives~\cite{huang2017chemotactic}.

Using the LBM-IBM implementation, we do not solve exactly the Stokes equation but need to retain the inertia term for time marching. Namely, we solve the Navier-Stokes equation in the limit of low Reynolds ($\Re$) number to approximate the Stokes equation. Thus, the actual non-dimensional governing equations solved by the LBM are 
\begin{subequations}\label{eq:ge_SI}
\begin{align}
    \label{eq:continuity}
    \grad \cdot \bu & = 0, \\ 
    \label{eq:momentum}
    \Re \lp \frac{\partial \bu}{\partial t} + \bu \cdot \grad \bu \rp & = -\grad p + \grad^2 \bu, \\ 
    \frac{\partial c}{\partial t} + \bu \cdot \grad c & = \frac{1}{\Pe}\grad^2 c.\label{eq:adv-diff}
\end{align}
\end{subequations}
We choose $\Re=0.5$ throughout the work, and the effect of inertia is reasonably weak as shown in Sec.~\ref{sec:valid} for a single disks and Sec.~\ref{sec:effec_inertia} for many.

The boundary conditions (BCs) for $c$ and $\bu$ at the surface of disks are: 
\begin{subequations}\label{eq:bc_SI}
\begin{align}
    \bn \cdot \grad c & = -1,\label{eq:bc_c_SI} \\
    \bu & = \bu_{\text{slip}} + \bU + \bOmega \times(\br - \bR), \label{eq:bc_u_SI} 
\end{align}
\end{subequations}
where $\bn$ is the outward unit normal vector,  $\bu_{\text{slip}} = \lp \bI -\bn\bn \rp \cdot \grad c$ is the slip velocity at the disk surface, $\br$ and $\bR$ denote the coordinates of
a specific point at the disk surface and the disk center, respectively. Here, $\bU = U\be_x + V\be_y$ and $\bOmega =\Omega \be_z$ represent the translational and rotational velocities of the disk, respectively.

Besides the flow, we need to solve for the time-dependent position of disks. Assuming disks and the surrounding liquid have the same densities, we update $\bU$ and $\bOmega$ of a disk using Newton's law of motion,
\begin{subequations}\label{eq:disk_ODE}
    \begin{align}
     \pi \Re \frac{\d \bU}{\d t} &= \bF,\\
    \frac{1}{2}\pi \Re \frac{\d \bOmega}{\d t} &= \bL,
    \end{align}
\end{subequations} 
where $\bF$ and $\bL$ denote the hydrodynamic force and torque on the disk. \eqnrefS{eq:disk_ODE} recovers to the force-free $\bF = \mathbf{0}$ and torque-free $\bL = \mathbf{0}$ conditions at $\Re=0$ as typically adopted to calculate the velocities of Stokesian swimmers~\cite{lauga2009hydrodynamics}. 

\begin{suppfigure}[tbh!]
    \centering
    \includegraphics[width=0.75\linewidth]{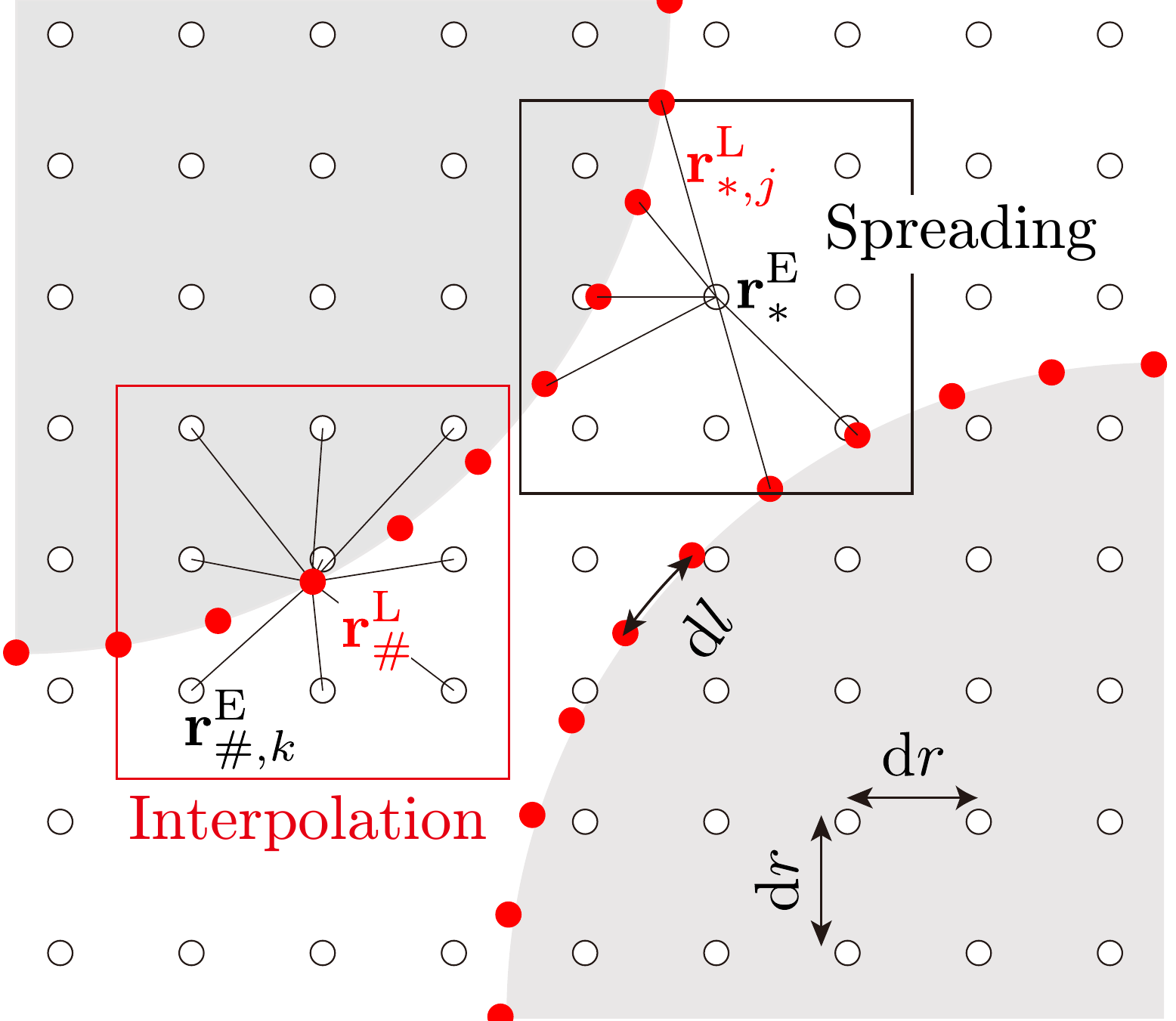}
    \caption{Schematic of the IBM implementation. The shaded areas indicate immersed disks. Hollow and filled circles mark Lagrangian and Eulerian grid points, respectively, with their spacings denoted by $\d l$ and $\dr$. Here, $\bre_{\ast}$ specifies a particular Eulerian point, and $\brl_{\ast,j}$ refers to its $j$-th (out of $6$ total here) neighbouring Lagrangian point. Likewise, $\brl_{\#}$ identifies a specific Lagrangian point, and $\bre_{\#,k}$ represents its $k$-th (out of $9$ in total here) neighbouring Eulerian point. The square boxes illustrate the support region of a three-point Dirac delta function used for interpolation and spreading in the IBM. }
    \label{fig:IBM_sketch}
\end{suppfigure}

\subsection{Lattice Boltzmann Method and Immersed Boundary Method}
We provide a brief introduction of the LBM-IBM solver, which is detailed in Refs.~\cite{jiang2019boundary,jiang2022simple}. We use a standard D2Q9 LBM with a single relaxation time to solve for the velocity field $\bu$ and solute concentration $c$. Besides the distribution functions for the  fluid density and velocity, we introduce another set of distribution functions $ \left\{ f_1,...,f_9 \right\}$ for $c$.

As shown in \figrefS{fig:IBM_sketch}, we adopt a uniform lattice grid (hollow circles) to discretize the Eulerian fluid domain and a uniform distribution of Lagrangian points (solid circles) to represent the disk edges. The spacings of the two grids are $\dr=1/\nrad$ and $\d l \sim O(\dr)$, respectively, where $\nrad$ is the number of lattices spanned by the disk radius. We choose $\nrad=15$ throughout this work following a grid dependence study detailed in Sec.~\ref{sec:valid}.

For LBM, the relaxation time $\sqrt{3}\nrad\Ma/\Re+0.5$ for the fluid motion determines the time step $\d t$. Here, $\Ma=\sqrt{3}\nrad\d t$ indicates the artificial compressibility of the LBM implementation, which should vanish in the Stokes flow. We choose $\Ma=\sqrt{3}/750$ that reconciles acceptable computational costs and respecting the $\Ma=0$ limit of interest.

\subsection{Immersed Boundary Method (IBM)} 
\subsubsection{IBM for the momentum equation} 
\label{subsec:momentum}
We employ a direct-forcing IBM~\cite{jiang2019boundary,jiang2022simple} to impose the velocity BC \eqnrefS{eq:bc_u_SI} at the disk surface. The presence of the immerse boundary is accounted numerically by introducing a source term $\bmF(\bre, t)$ into the momentum equation \eqnrefS{eq:momentum}. The modified equation takes the form of
\begin{align} \label{eq:dic-momentum}
    \nonumber \Re \left[ \frac{\partial \bu \lp \bre, t \rp}{\partial t} + \bu \lp \bre, t \rp \cdot \grad \bu \lp \bre, t \rp \right] & =  \\   
    -\grad p \lp \bre, t \rp + \grad^2 \bu \lp \bre, t \rp + \bmF \lp \bre, t \rp. 
\end{align}

We now explain how to calculate the source term $\bmF$ at a specific Eulerian point $\bre_*$, see \figrefS{fig:IBM_sketch}. 
We first identify its neighbouring Lagrangian points $\brl_{*,j}$ ($j$ is the index of the neighbours) within the compact support of a three-point Dirac delta function $D(\bre_*-\brl_*)$. Here, $D(\hbr)$ is defined by
\begin{align}
    D(\hbr)  
    = 
     \frac{1}{\dr^2} \dtp\lp \frac{|\hbr \cdot \be_x|} {\dr} \rp \dtp \lp \frac{|\hbr \cdot \be_y|} {\dr} \rp, 
\end{align}
where $\dtp(r)$ is a three-point regularized delta function~\cite{jiang2019boundary}, \begin{align*}
    \dtp(r) = 
    \Biggl\{
    \begin{array}{ll}
        0,  & r>1.5, \\
        \frac{1}{6}(5-3r-\sqrt{1-3(1-r)^2}),  &0.5 < r \leq 1.5, \\
        \frac{1}{3}(1+\sqrt{1-3r^2}),  & r \leq 0.5.
    \end{array}
\end{align*}

For the Eulerian point $\bre_*$, we calculate an intermediate velocity at all its neighboring Lagrangian points. For example, at a particular Lagrangian point $\brl_{\#}$, we can obtain its intermediate velocity $\bup(\brl_{\#,j},t)$ via interpolation,
\begin{align} \label{eq:u_SI}
  \bup(\brl_{\#},t)=\sum_k \bup (\bre_{\#,k},t)D\lp \bre_{\#,k}-\brl_{\#} \rp \d r^2,
\end{align}
where $\bre_{\#,k}$ denotes the $k$-th (out of all) neighbouring Eulerian point of $\brl_{\#}$.

Having known $\bup(\brl_{*,j},t)$, we compute the force per unit volume $\bf(\brl_{*,j},t)$ acting on the Lagrangian point. It is determined, 
at each time step,  by considering the difference between the velocity at $\brl_{*,j}$ prescribed by \eqnrefS{eq:bc_u_SI} and the intermediate one,
\begin{align}\label{eq:IB_force_L}
    \bf(\brl_{*,j},t) = \frac{2 \Re }{\d t}[ \bu(\brl_{*,j},t) - \bup(\brl_{*,j},t) ].
\end{align}  

The total hydrodynamic force $\bF$ and torque $\bL$ acting on the disk are then determined based on the force density of all its Lagrangian points,
\begin{subequations}\label{eq:FL_Disk_SI}
    \begin{align}
        \bF & = \sum_j \bf(\brl_{*,j},t)\d l \dr,\label{eq:F_Disk_SI}\\   
        \bL & = \sum_j (\brl_{*,j} - \bR) \times \bf(\brl_{*,j},t) \d l \dr. \label{eq:L_Disk_SI} 
    \end{align}
\end{subequations}
Subsequently, we employ the forward Eulerian scheme to solve \eqnrefS{eq:disk_ODE} and update the translational $\bU$ and rotational $\bOmega$ velocities of the disk by using $\bF$ and $\bL$, respectively.

Finally, at the Eulerian point $\bre_*$, we determine $\bmF(\bre_*,t)$ 
by spreading $\bf(\brl_{*,j},t)$ from its neighboring Lagrangian points 
\begin{align}
    \bmF(\bre_*,t) = -\sum_j \bf(\brl_{*,j},t)D\lp \brl_{*,j}-\bre_* \rp \d l \dr.
\end{align}

\subsubsection{IBM for the advection-diffusion equation}
Besides the velocity, a direct-forcing IBM is also applied for the flux BC \eqnrefS{eq:bc_c_SI}. It is achieved through another direct-forcing IBM specifically designed for thermal problems~\cite{Ren2012boundary}. The more general form of \eqnrefS{eq:bc_c_SI} reads
\begin{align}\label{eq:bc_2_SI}
 \bn \cdot \grad c(\brl,t)=-m(\brl,t) \Pe /\dr,
\end{align}
which includes \eqnrefS{eq:bc_c_SI} as a special case of $m=\dr/\Pe$. The essential idea for imposing a flux on immersed boundaries is to determine, at every time step, a source term $M(\bre,t)$ introduced in \eqnrefS{eq:adv-diff}. The modified equation is
\begin{align} \label{eq:dic-adv-diff}
   &\frac{\partial c\lp \bre, t \rp}{\partial t} + \bu \lp \bre, t \rp \cdot \grad c\lp \bre, t \rp \nonumber \\
   &= \frac{\grad^2 c\lp \bre, t \rp}{\Pe} + M \lp \bre, t \rp.
\end{align}
Calculating $c(\bre,t)$ at Eulerian points starts with an intermediate prediction $\cp(\bre,t)= \sum_i f_i(\bre,t)$ without satisfying the BC. Then, the predicted value is compensated by a correction 
$\cc(\bre,t)=M(\bre,t) \d t/2$ proportional to the source term \cite{kang2011comparative}, namely $c(\bre,t)=\cp(\bre,t)+\cc(\bre,t)$.

We now illustrate the procedure for calculating the source term $M$ at a specific Eulerian point $\bre_*$. After identifying its neighboring Lagrangian points $\brl_{*,j}$, we compute the corresponding intermediate concentration flux

\begin{align} \label{eq:inter_flux}
    \mp(\brl_{*,j},t) = -\bn \cdot \grad \cp(\brl_{*,j},t) \d r/\Pe,
\end{align}
where $\grad \cp(\brl_{*,j},t)$ is the intermediate concentration gradient at  $\brl_{*,j}$. Likewise, at a particular Lagrangian point $\brl_{\#}$, we can obtain this gradient  $\grad \cp$ via interpolation,
\begin{align} \label{eq:grad_c_SI}
 \grad \cp(\brl_{\#},t)=\sum_k\grad \cp(\bre_{\#,k},t)D\lp \bre_{\#,k}-\brl_{\#} \rp \d r^2.
\end{align}

Upon acquiring $\mp(\brl_{*,j},t)$, we compute the difference
\begin{align*}
    \d m(\brl_{*,j},t) = m(\brl_{*,j},t) - \mp(\brl_{*,j},t)
\end{align*}  
between the expected value $m(\brl_{*,j},t)$ and the intermediate one. Finally, we determine $M(\bre_*,t)$ 
by spreading the flux differences from its neighboring Lagrangian points
\begin{align}
    M(\bre_*,t) = \sum_j 2\d m(\brl_{*,j},t)D\lp \brl_{*,j}-\bre_* \rp \d l.
\end{align}

\subsection{Validation of The LBM-IBM Solver}\label{sec:valid}
\subsubsection{Hydrodynamics alone: A 2D squirmer swimming in a closed box}\label{sec:squirmer}
\begin{suppfigure}[tbh!]
    \centering
    \includegraphics[width=1.0\linewidth]{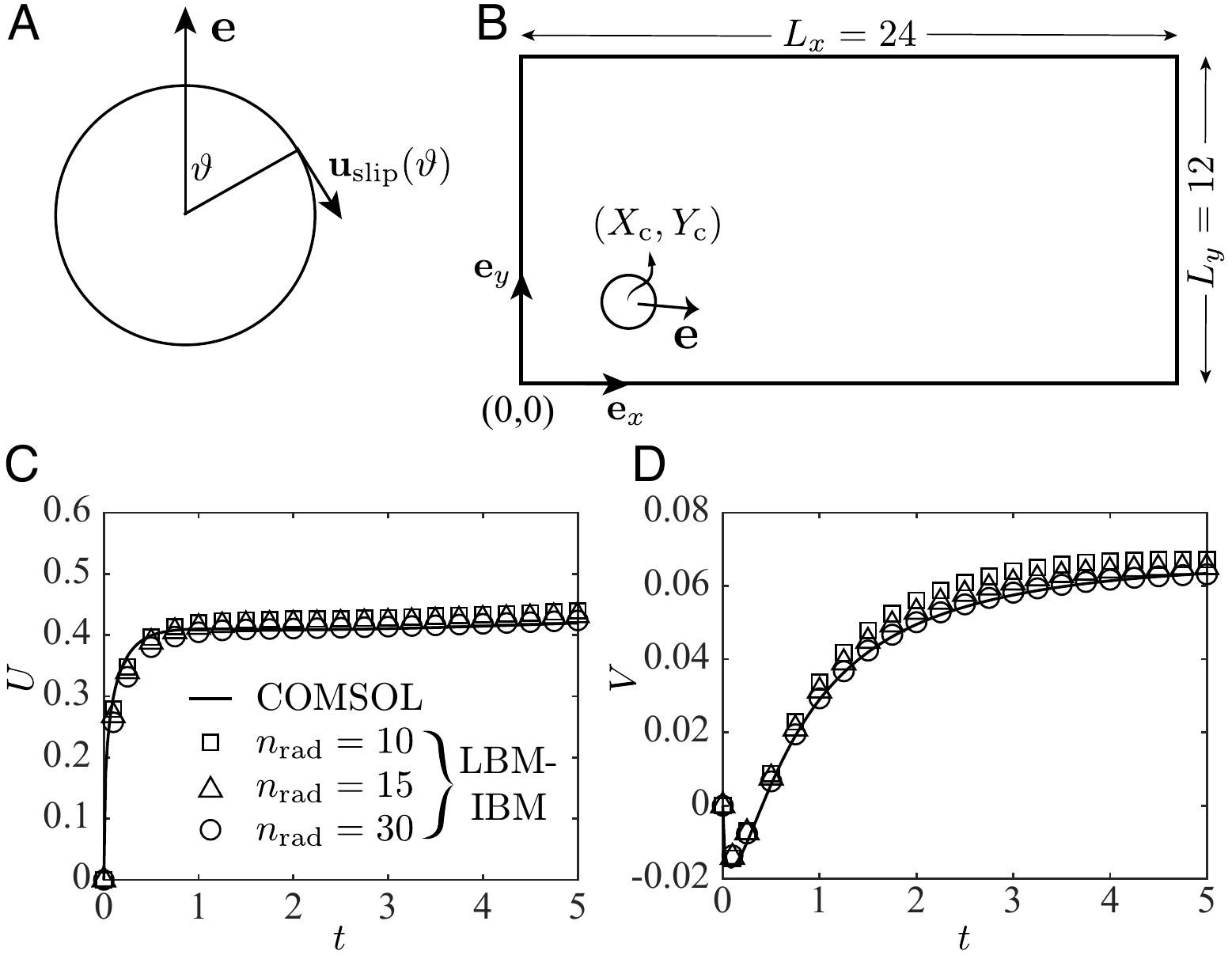}
    \caption{
    A, A 2D squirmer self-propels by its surface actuation represented by a slip velocity $\buslip\lp\thetas \rp$ symmetric about its orientation $\be$. Here, $\thetas$ is the polar angle at the surface with respect to $\be$. B, Locomotion of a squirmer in a closed box with length $L_x=24$ and height $L_y=12$. The squirmer centered at $(\Xc,\Yc)$ swims at a velocity $\bU = U\be_x + V\be_y$. C and D, Velocity components $U$ and $V$ evolve over time $t$; the curve and symbols correspond to the COMSOL and LBM-IBM data, respectively. 
    }
    \label{fig:squimer}
\end{suppfigure}

Focusing on a single active particle, we validate our LBM-IBM solver against theory and/or numerical data obtained via finite element method (FEM). The FEM implemented in COMSOL Multiphysics (I-Math, Singapore) uses a body-fitted mesh other than the diffused interface of IBM, which thus guarantees the accuracy of benchmark solutions. The validation is performed for two benchmark settings: 1) a model microswimmer with prescribed surface actuation swimming in a box, and 2) the spontaneous propulsion of an isotropic phoretic disk; the former involves hydrodynamics alone and the latter features coupled hydrodynamics and solute transport.

We consider the locomotion of a circular model microswimmer moving in a 2D closed box. Our model swimmer, the widely studied squirmer, was initially proposed to mimic ciliated microorganisms~\cite{blake1971spherical,lighthill1952squirming}. The 2D squirmer self-propels using a surface actuation symmetric about its orientation $\be$. This actuation is represented by a tangential slip velocity $\buslip(\thetas)$ distributed along  the polar angle $\thetas$ with respect to $\be$, as shown in \figrefS{fig:squimer}{A}. We adopt a common velocity distribution 
\begin{align}\label{eq:sq_bc}
\buslip=\lp \sin\thetas + \beta
\sin \thetas\cos \thetas\rp  \be_{\thetas}, 
\end{align}
including the first two squirming modes~\cite{blake1971spherical}. Here, $\beta$ indicates the ratio of the second mode to the first mode. We choose the first mode as the characteristic velocity  rather than $\mA \mM /\mD$ of the hydrochemical problem, leaving the resulting dimensionless equations for fluid motion  unchanged.

Using COMSOL and LBM-IBM, we solve Eqs.~\eqref{eq:continuity}, \eqref{eq:momentum} and~\eqref{eq:disk_ODE} for the setting depicted in \figrefS{fig:squimer}{B}.
The box's length and height are $L_x=24$ and $L_y=12$, respectively, with the origin $(x,y)=(0,0)$ at its left-bottom corner. No-slip BCs are imposed on all  four sides. At $t=0$, the squirmer, centered at $\lp \Xc,\Yc \rp = \lp 3,4 \rp$, is oriented slightly towards the bottom wall; the angle between its orientation $\be$ and the $\be_x$ axis is $5^{\circ}$. We choose $\beta=2$ and $\Re=0.5$ for this study. 

We first obtain the benchmark solution from COMSOL. Approximate $10^5$ Taylor-Hood elements are used for discretization, allowing us to eliminate the unnecessary streamwise diffusion invoked by default. The mesh is refined near the disk, with the smallest and largest elements sized $0.025$ and $0.25$, respectively. In the frame of COMSOL, we adopt the Global Equations node to solve \eqnrefS{eq:disk_ODE} and the Moving Mesh Interface to handle the time-evolving domain.

We conduct LBM-IBM simulations with three grid resolutions featuring $\nrad =10, 20$, and $30$. 
Increasing this resolution yields better agreement between the LBM-IBM and COMSOL data (see \figrefS{fig:squimer}{C-D}). 
Their excellent matching when $\nrad=30$ suggests that both solvers have been cross-validated, considering their completely different numerical algorithms. 
Notably, our COMSOL implementation for diverse chemically phoretic swimmers, including disks and particles, has undergone thorough validation, as detailed in Ref.~\cite{zhu2023self}.

\subsubsection{Coupled hydrodynamics and solute transport: Spontaneous motion of a phoretic disk} 
\begin{suppfigure}[tbh!]
    \centering
    \includegraphics[width=0.7\linewidth]{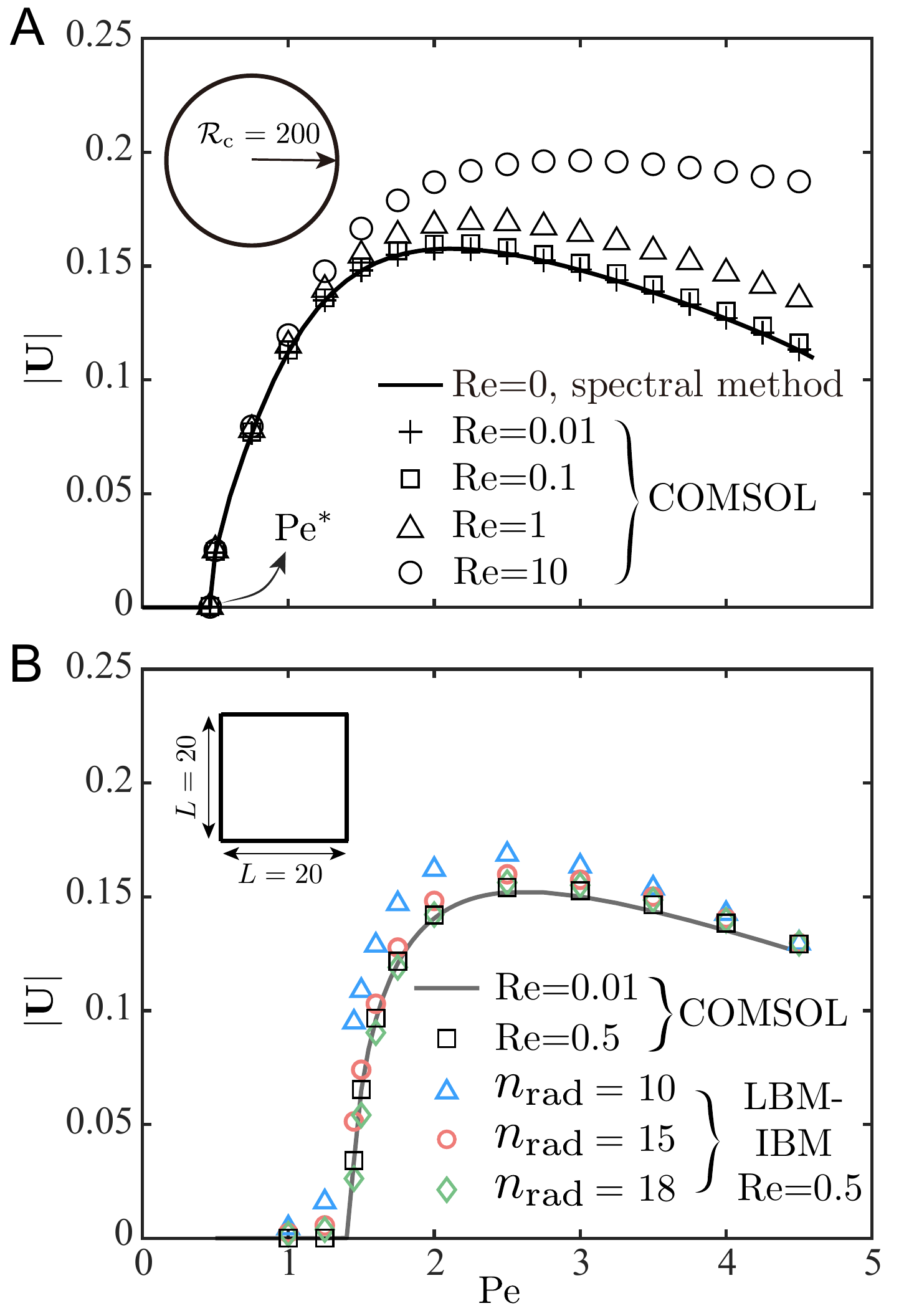}
    \caption{
    A two-step approach to validate the LBM-IBM implementation. A, firstly, developing a COMSOL solver firmly validated against a spectral code~\cite{hu2019chaotic} for a spontaneously swimming phoretic disk in a circular domain of radius $\Rc=200$. B, secondly, validating the LBM-IBM approach against the COMSOL solver developed in the first step for a spontaneously swimming phoretic disk in a square domain sized $L=20$. More details can be found in the text.
    }
    \label{fig:phoretic_hu}
\end{suppfigure}

Having validated our LBM-IBM solver for the pure hydrodynamic problem, we  turn to the hydrochemical scenario of an isotropic phoretic disk, mathematically described in Sec.~\ref{sec:equation_SI}. For a single disk centered in a circular fluid domain, the critical P\'eclet number $\Pecri$ corresponding to the onset of instability has been derived theoretically by Ref.~\cite{hu2019chaotic} in the creeping flow regime, \viz $\Re=0$ (inertialess limit); the authors also used a Chebyshev spectral method to calculate the disk's swimming velocity for various unstable $\Pe$ values. A Dirichlet BC $c=0$ is imposed at the outer boundary of the domain.

Considering the inconvenience of discretizing a curved domain with LBM, we adopt a two-step approach to validate our LBM-IBM solver: first, we develop a COMSOL implementation convincingly validated against Ref.~\cite{hu2019chaotic} for a disk in a circular domain of radius $\Rc$ (see \figrefS{fig:phoretic_hu}{A}); second, we use this FEM solver to benchmark the LBM-IBM counterpart for a disk in a square domain sized $L$. Here, we exploit the flexibility of FEM in discretizing complex domains and its intact validity subject to domain variation.

In the first step using COMSOL alone, we solve the flow in the frame of the disk. Hence, Dirichlet BCs $\bu = -\bU$ and $c=0$ are specified at the outer edge of the domain. The undeformed ring-shaped fluid domain is discretized by approximately $32000$ triangular elements. Besides the Taylor-Hood discretization for the flow equation as in Sec.~\ref{sec:squirmer}, the elements are of second order to represent the concentration $c$. The mesh is refined near the disk and enlarged in the far field, with sizes of the smallest and largest elements being $0.05$ and $9.24$, respectively. We choose a specific domain size $\Rc=200$  to compare our results with those of Ref.~\cite{hu2019chaotic}. In addition to the inertialess limit $\Re=0$~\cite{hu2019chaotic}, we vary $\Re \in [0.01, 10]$  to probe the effect of inertia as a by-product of the validation. Moreover, $\Pe$ is limited to the range when the disk is stationary or swimming steadily. As shown in \figrefS{fig:phoretic_hu}{A}, the disk's swimming speed $|\bU|$ at different $\Re$ from COMSOL simulations agrees well with the prediction of Ref.~\cite{hu2019chaotic} near the onset of instability. Inertia plays a negligible role in this regime. Additionally, COMSOL results lie on top of the benchmark solution ($\Re=0$) for all $\Pe$ values when $\Re=0.01$. The speed is observed to increase with $\Re$, becoming more pronounced at larger $\Pe$. When $\Pe=4.5$, the speed at $\Re=10$ is larger than its inertialess limit by about $65\%$.

Now we use the validated COMSOL solver to benchmark our LBM-IBM implementation, targeting a phoretic disk in a square domain of size $L=20$ (\figrefS{fig:phoretic_hu}{B}). The COMSOL setup differs from that in the first step (\figrefS{fig:phoretic_hu}{A}) only in the shape and size of the fluid domain, thus the equations are also solved in the body frame with unchanged BCs. On the other hand, the LBM-IBM simulations are conducted in the lab frame, with the corresponding BCs $\bu=\mathbf{0}$ and $c=0$ imposed at the four sides of the domain. In COMSOL, the disk remains in the domain center and is equidistant from the four boundaries. For a reliable comparison, the same effect of boundaries is realized in LBM based on the moving-grid technique~\cite{nie2020simulation}. It is important to note that this technique is used only for validation purposes here but not for the cases of suspensions in a periodic domain. Using COMSOL, we calculate the $\Pe$-dependent swimming speed $|\bU|$ for $\Re=0.01$ and $\Re=0.5$. The small difference between the two datasets suggests a reasonably weak inertia effect at $\Re=0.5$, which can be chosen to approximate the Stokes flow of interest. Subsequently, $|\bU|\lp \Pe \rp$ at $\Re=0.5$ is computed with LBM-IBM using three grid resolutions $\nrad=10$, $15$, and $18$. As expected, increasing the resolution $\nrad$ produces better agreement between the COMSOL and LBM-IBM data. Based on this agreement, we believe that a combination of  $\Re=0.5$ and $\nrad=15$ achieves an appropriate balance between computational cost and precision for simulating Stokesian suspensions of active microswimmers. A reduced $\Re$ and increased resolution will be adopted in our future studies. 

\begin{suppfigure}[tbh!]
    \centering
    \includegraphics[width=0.6\linewidth]{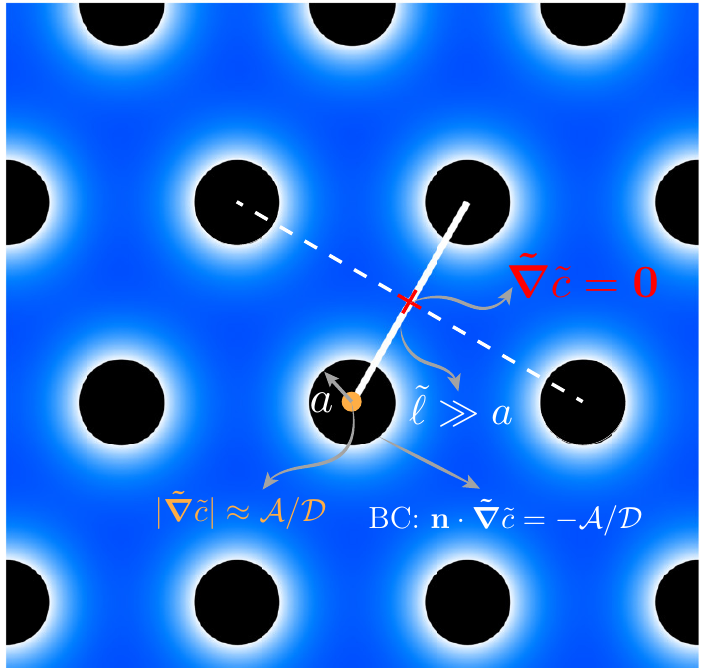}
    \caption{Schematic for better explaining the scaling theory. Black circles denote the disks forming a hexagonal lattice. The text provides more details.}
    \label{fig:stability}
\end{suppfigure}

\begin{suppfigure*}[tbh!]
\centering
\includegraphics[width=0.75\linewidth]{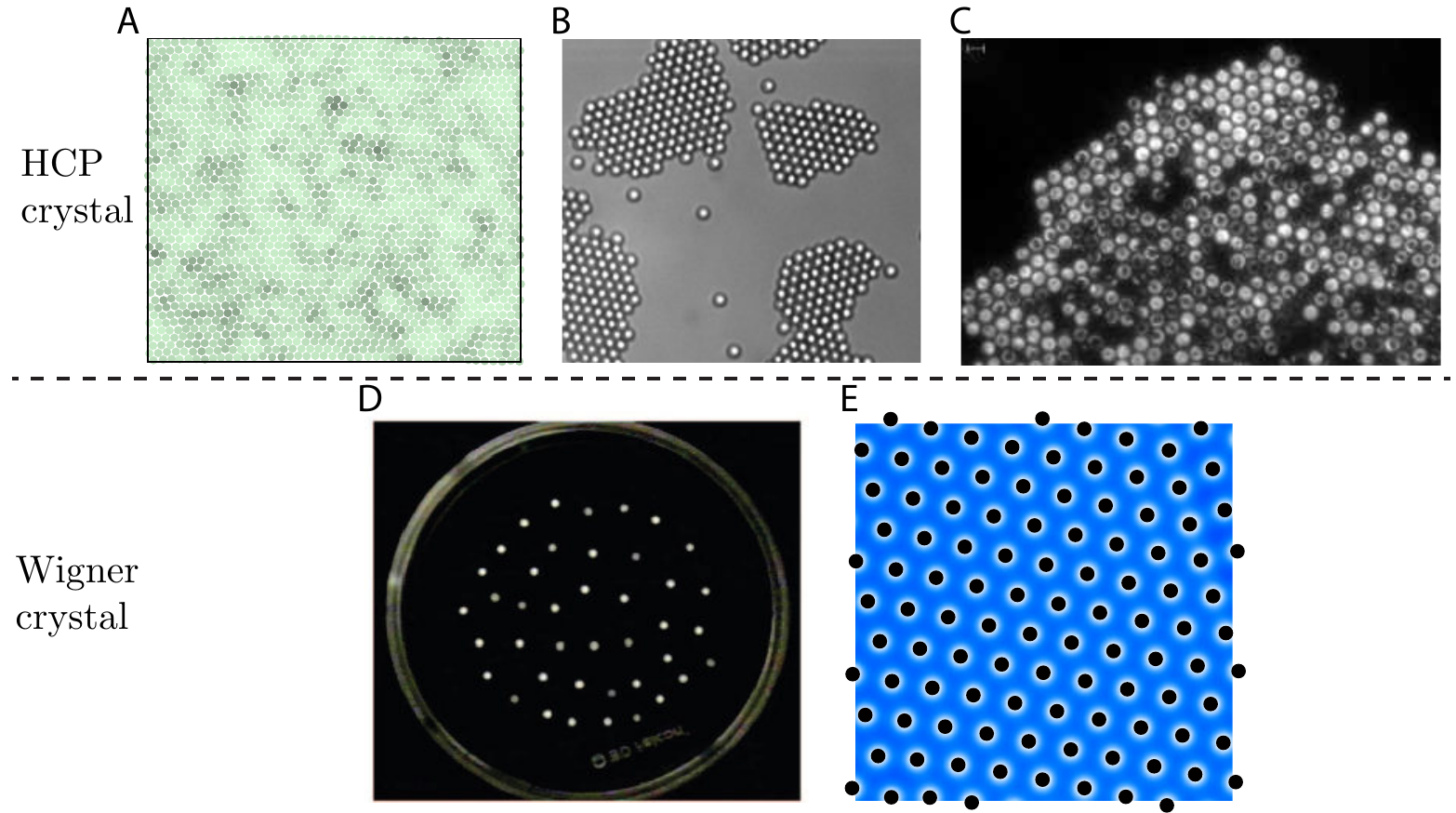}
\caption{Hexagonal closed-packed crystal (upper row) and hexagonal Wigner crystal (lower row) formed in active matter systems. A, Brownian dynamics computer simulations of a Yukawa model of self-propelled particles, adapted with permission from~\cite{bialke2012crystallization}. Copyright (2012) by the American Physical Society. B, a monolayer of photoactivated Janus colloids, adapted from~\cite{palacci2013living} with permission from AAAS. C, bacterial crystal formed by \textit{Thiovulum majus}, adapted with permission from~\cite{petroff2015fast}. Copyright (2015) by the American Physical Society. 
D, Wigner crystal-like lattice formed by camphor surfers, reproduced with permission from~\cite{soh2008dynamic}. Copyright 2008 American Chemical Society. E, Wigner crystal of isotropic phoretic agents we observe (Fig. 1\textbf{d} of the main article), with an area fraction of $\phi=0.12$. 
}
\label{fig:crystal_diff}
\end{suppfigure*}

\section{Supplementary Discussion}

\subsection{Theoretical scaling for the solid-liquid phase transition}\label{sec:theory}

In the main article (shown in Fig. 2), we develop a scaling to predict the transition from the solid to liquid phase. In the solid phase, disks self-organize into a stationary hexagonal lattice. 
This base state becomes unstable when the phoretic activity $\Pe$ surpasses a certain threshold. Accordingly, the disks start to swim, leading to the liquidation of the hexagonal state.

The hexagonal base state is portrayed in \figrefS{fig:stability} to  help rationalize the scaling argument. The argument relies on three aspects. First, realizing the BC $\bn\cdot \tgrad \tc  = -\mA/\mD$ at the disks surface and assuming a much larger inter-disk space $\tell$ than $a$ in the limit of low $\phi$, we can approximate $\tgrad \tc$ at the disk center by $\mA/\mD$. Second, $\tgrad \tc = \mathbf{0}$ in the middle (marked by a cross) of two neighbouring disks; indeed, the two mirror symmetries at the middle position imply that the gradients of $\tc$ in the two orthogonal directions are both strictly zero, leading to zero $\tgrad \tc$. Third, we have implicitly assumed an infinite-fold rotational symmetry (circular symmetry) that is stronger than the actual six-fold rotational symmetry, leading to zero $\tgrad \tc$ everywhere on the ring of radius $\tell/2$ centered at each disk.

\subsection{Active Wigner crystal}
We provide further discussion on the hexagonal crystal formed by the phoretic disks, as exemplified by 
Fig. 1\textbf{d} of the main article. Our identified hexagonal structure differs from the previously observed ones formed 
by other active swimmers~\cite{bialke2012crystallization, palacci2013living,petroff2015fast,briand2016crystallization, singh2016universal,klamser2018thermodynamic, kichatov2021crystallization}. 
Besides the different mechanisms for crystallization, an evident difference is the packing fraction (PF) of crystals.

In the above-mentioned works, the observed crystal presents a monolayer hexagonal close-packed (HCP) arrangement (see the upper row of \figrefS{fig:crystal_diff}); in this setting, the swimmers representing atoms pack as densely as possible, hence approaching the maximum possible PF of $\pi/\lp 2\sqrt{3} \rp \approx 0.907$~\cite{kittel2018introduction, west2022solid}---not to be confused with the well-known PF of $\pi/\lp 3\sqrt{2}\rp \approx 0.74$ for the more general three-dimensional HCP arrangement. A graphene layer typically features $\text{PF} \approx 0.907$, whereas metals like magnesium and titanium crystallize in the HCP structure with $\text{PF} \approx 0.74 $.

In contrast, the PF mirroring the area fraction $\phi$ we have defined---of our examined crystal structure can plummet to as low as $0.005$. This scenario corresponds to the data point at the lower left corner in Fig. 2 of our manuscript. Such a crystalline configuration embodies what is known as the Wigner crystal, exemplified by the lower row of \figrefS{fig:crystal_diff}).   Theoretically predicted by Eugene Wigner in 1934~\cite{wigner1934interaction, wigner1938effects}, this solid phase of electrons was only recently visualized in experiments~\cite{li2021imaging}. A Wigner crystal forms at low electron densities, where the distance between electrons significantly exceeds the effective ``electron size'', denoted by the wave function's extent. The separation of these two length scales enables long-range Coulomb repulsion between electrons to overpower their kinetic energy, thus giving rise to an ordered lattice structure.

The formation of a Wigner crystal from our phoretic disks can be traced back to the established equivalence between phoretic and Coulomb interactions~\cite{illien2017fuelled, golestanian2019phoretic, liebchen2021interactions}: both exhibit repulsive characteristics and their potential decays slowly as $1/\hat{r}$ with $\hat{r}$ being the distance. 
Drawing parallels with the conditions for the formation of the original Wigner crystal---where the electrons' kinetic energy is significantly weaker than their repulsion---in our case of phoretic agents, they crystallize when their activity, indicated by the Pe number, falls below a certain threshold.

In addition to the collective behaviour of phoretic disks, we have also examined the pairwise interaction between two disks. As shown in \figrefS{fig:trajectory}{A} (see \movRef{9}), our simulations reproduce the typical crossing and reflecting trajectories of two phoretic swimmers previously modelled~\cite{lippera2021alignment,hokmabad2022chemotactic} and experimentally observed on active droplets~\cite{hokmabad2022chemotactic}. Here, we simulate two or three disks in a periodic domain of $L=100$ and set $\Pe=2.5$, at which an isolated disk swims steadily. Beyond these two scenarios, we identify a stable bound state of two disks swimming in parallel, as illustrated in \figrefS{fig:trajectory}{B} and \movRef{10}. 

\subsection{Interaction between two phoretic disks}\label{sec:interaction}
\begin{suppfigure}[tbh!]
    \centering
    \includegraphics[width=1\linewidth]{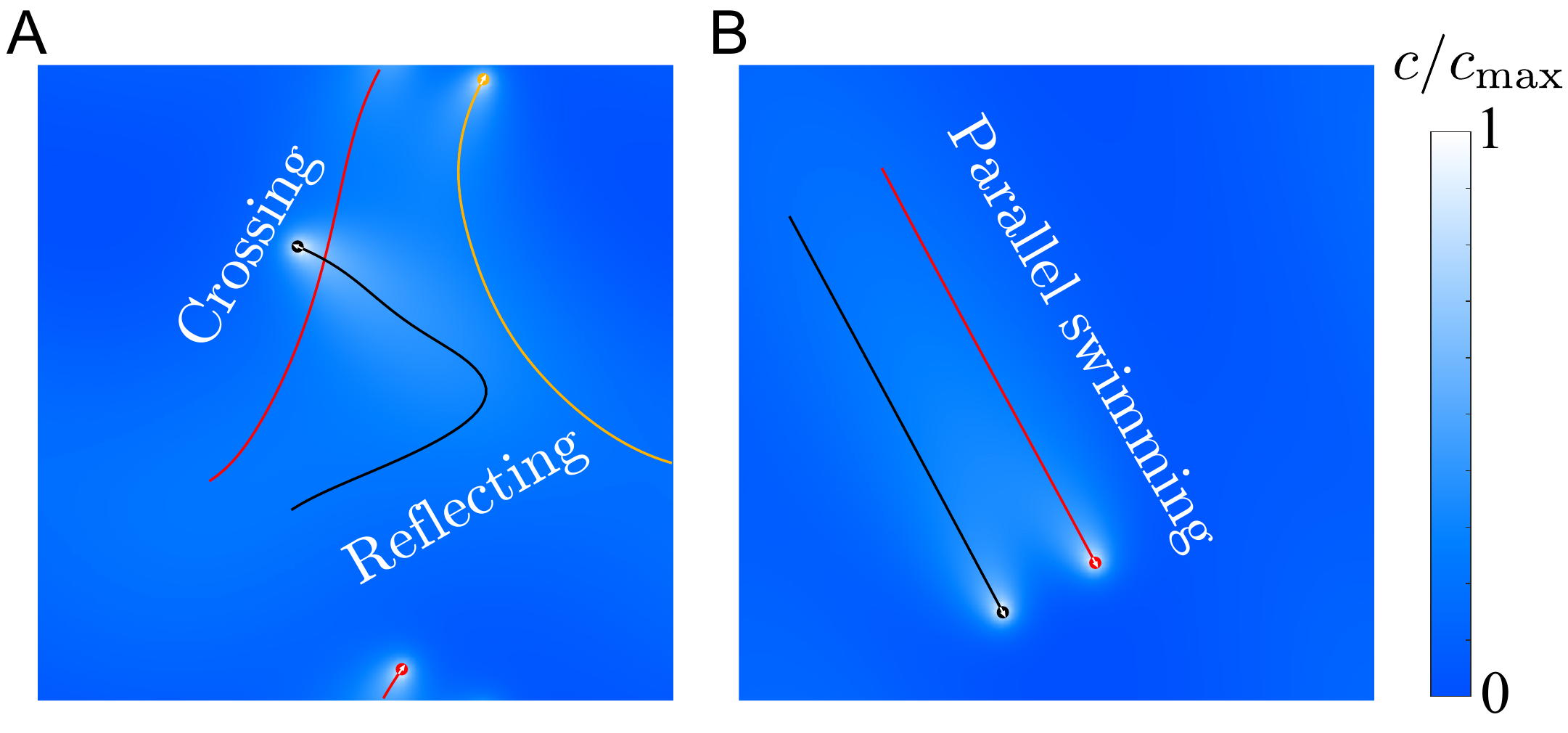}
    \caption{Characteristic scenarios of two interacting phoretic disks at $\Pe=2.5$ in a periodic domain of $L=100$. A, crossing or reflecting trajectories of two disks. B, two disks form a stable bound pair swimming in parallel. The colormap shows the scaled concentration $c/\cmax$.}
    \label{fig:trajectory}
\end{suppfigure}

\begin{suppfigure*}[tbh!]
    \centering
    \includegraphics[width=0.65\linewidth]{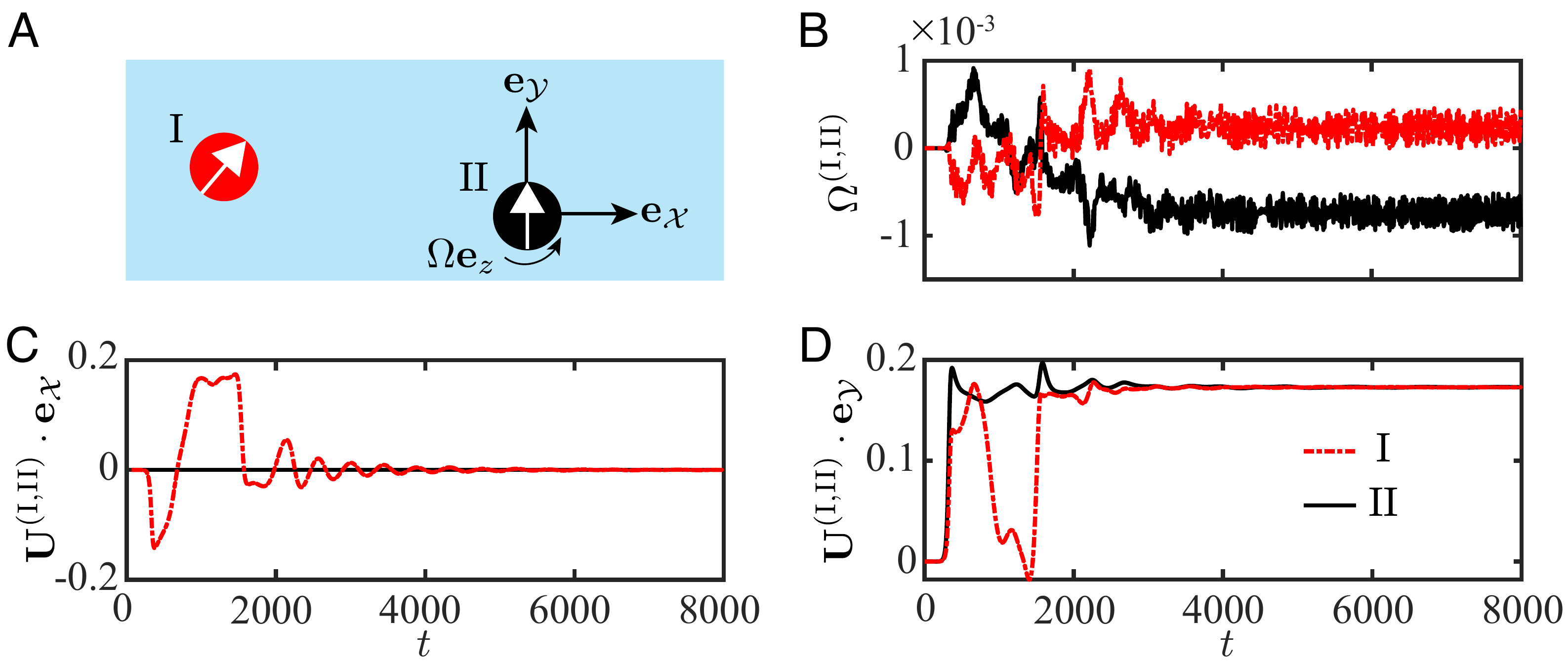}
    \caption{
    A, schematic of two interacting disks numbered I (red) and II (black). B, time evolution of their rotational velocities $\Omega^{\lp \text{I}\rp}$ and $\Omega^{\lp \text{II}\rp}$.  C and D, similar to B, but for the velocity components $\bU^{\lp \text{I},\text{II} \rp}\cdot \epen$ and $\bU^{\lp \text{I},\text{II} \rp}\cdot \epar$ in the local frame $\be_{\mathcal{X}\mathcal{Y}}$, respectively.
    }
    \label{fig:parallel_vel}
\end{suppfigure*}

\begin{suppfigure*}[tbh!]
    \centering
    \includegraphics[width=0.9\linewidth]{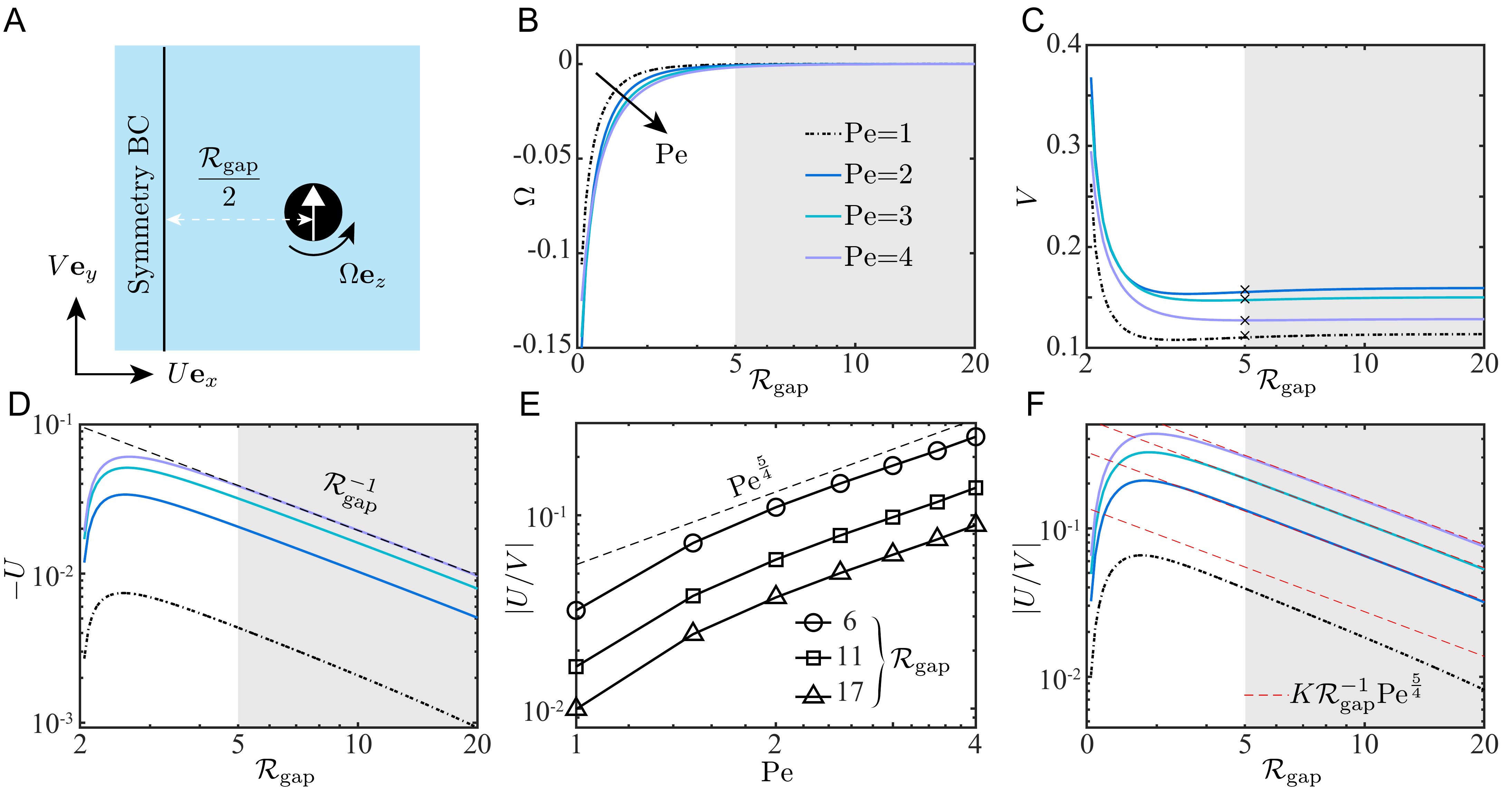}
    \caption{
A, numerical setup for examining the HI among two swimming phoretic disks separated by $\Rgap$. A symmetry BC is used here. The text provides detailed information on the setup. B-D, the disk's rotational velocity $\Omega$, and translational velocity components $V$ and $-U$, respectively, versus the inter-disk distance $\Rgap$ at varying $\Pe\in[1,4]$. The shaded area indicates the regime in which the results are considered consistent with the adopted assumption of negligible disk rotation. The crosses in C represent the swimming speed of the corresponding isolated disks. A log-log scale is adopted in D. E, $|U/V|$ quantifying the HI-induced attraction between disks versus $\Pe$ at varying distances. Numerical data (symbols) are compared to the fitted curves. F, fitting $|U/V|\lp \Pe, \Rgap \rp $ by $K \Rgap^{-1}\Pe^{5/4} $ with $K\approx 1.5$.
}
    \label{fig:parallel_twop_comsol}
\end{suppfigure*}
The crossing and reflecting scenarios can be captured by models without considering hydrodynamics. However, the stable parallel state has only been discovered in experiments naturally involving hydrodynamics and hydrodynamic interactions (HIs), as reported very recently~\cite{hokmabad2022spontaneously}. In contrast, it has not been captured by modelling before. Considering this difference and the state of two bound swimmers reproduced by our hydrodynamic simulations, we speculate that this emerging state is related to their HI, as analyzed below.

We scrutinize the velocity evolution of the two disks numbered I (red trajectory) and II (black trajectory) in \figrefS{fig:trajectory}{B}. To focus on their relative motion, we define a local frame $\be_{\mathcal{XY}}$ attached to the center of disk II, as shown in \figrefS{fig:parallel_vel}{A}. Here, $\epar$ is aligned with its swimming direction, and rotating $\epar$ clockwise by $90^{\circ}$ results in $\epen$. We characterize the emergence of the parallel bound swimming state by the time evolution of the disks' rotational velocity velocities $\Omega$, as well as their local velocity components $\bU^{\lp \text{I,II}\rp} \cdot \epen$ and $\bU^{\lp \text{I,II}\rp} \cdot \epar$. After entering their swimming states, the two swimmers initially depart ($\bU^{\lp \text{I}\rp} \cdot \epen<0$) and then attract each other ($\bU^{\lp \text{I} \rp } \cdot \epen>0$), as indicated by \figrefS{fig:parallel_vel}{C}. Following periodic switches between the two states with a decaying strength, the swimmers eventually form a parallel bound pair with the swimming velocity $\bU \cdot \epar$ (see \figrefS{fig:parallel_vel}{D}). Meanwhile, the two disks rotate in opposite directions at a magnitude negligible compared to their translational speed (see \figrefS{fig:parallel_vel}{B}).

The departing behaviour ($\bU^{\lp \text{I}\rp} \cdot \epen<0$ in \figrefS{fig:parallel_vel}{C}) of the two disk swimmers is known to result from their chemo-repulsive interaction~\cite{lippera2020collisions,hokmabad2022chemotactic}. On the other hand, their attraction ($\bU^{\lp \text{I}\rp} \cdot \epen>0$) is attributed to the inter-swimmer HI, as evidenced experimentally for chemically active droplets~\cite{hokmabad2022spontaneously}. 
These droplet swimmers are recognized to behave like a pusher-type squirmer~\cite{michelin2013spontaneous,hokmabad2022chemotactic} (see the left panel of Fig. 1\textbf{c} of the main article), exhibiting a dipolar flow pattern that pulls fluid in from the sides~\cite{zottl2016emergent}. Consequently, two swimmers will experience hydrodynamic attraction towards each other from the sides.

Here, we devise a two-step numerical approach that decouples hydrodynamics from phoretic dynamics, allowing us to focus on the role of HI resulting from the dipolar flow pattern. The simulations are conducted using the COMSOL implementation described in Sec.~\ref{sec:valid}. First, we extract the slip velocity at the surface of a steadily swimming phoretic disk, where we solve the hydrochemical problem; the slip velocity is left-right symmetric about its swimming orientation. Second, we study two identical close-by disks propelled by the slip velocity extracted from the first step when addressing the hydrodynamic problem alone. Realizing the negligible disk rotation within the bound state (see \figrefS{fig:parallel_vel}{B}), we consider an instantaneous configuration of two parallel-oriented squirmer-like disks, whose connecting line is normal to their common orientation (see   \figrefS{fig:parallel_twop_comsol}{A}). Exploiting the mirror symmetry of this setting, we simulate one swimming disk using a symmetry BC. 

\subsection{Formation of arc-shaped disk chains}\label{sec:chain}
\begin{suppfigure}[tbh!]
    \centering
    \includegraphics[width=1\linewidth]{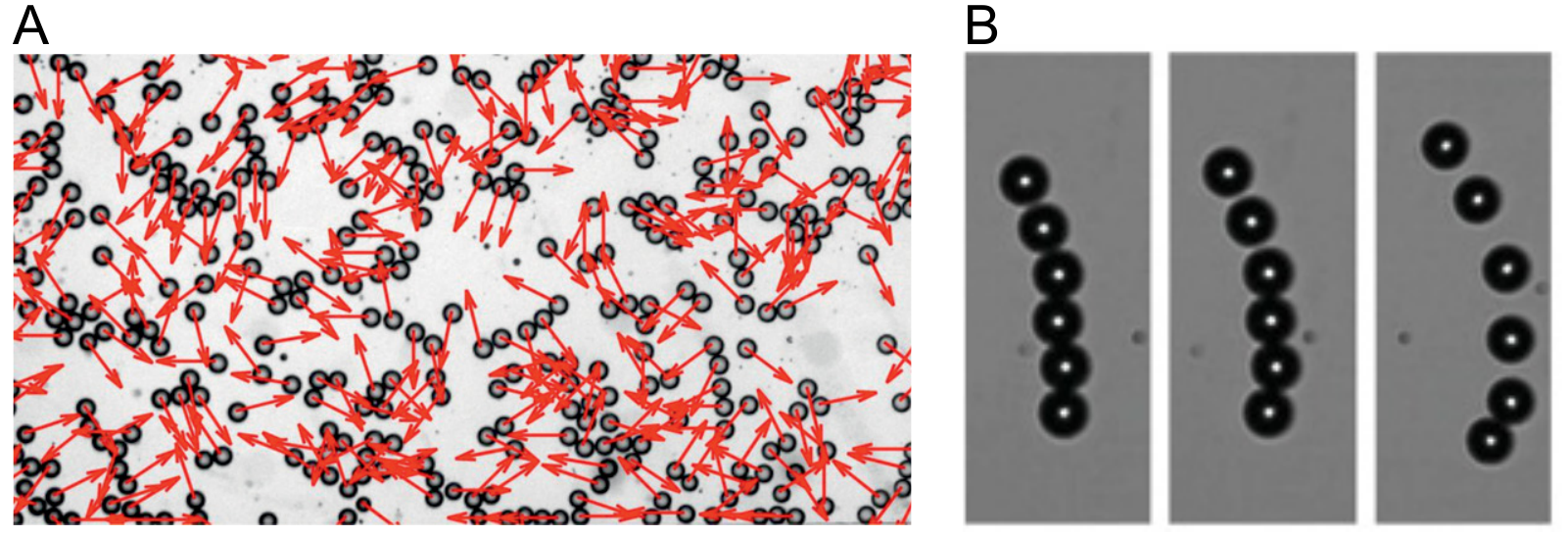}
    \caption{Chain formation of chemically active droplets observed in experiments. A is reproduced from  Ref.~\cite{thutupalli2011swarming} by permission of IOP Publishing under the  CC BY-NC-SA licence.     \raisebox{.5pt}{\textcircled{\raisebox{-.9pt} {C}}} Deutsche Physikalische Gesellschaft. 
    B is reproduced from Ref.~\cite{thutupalli2018flow} with permission from the National Academy of Sciences (NAS). 
}
    \label{fig:chain_exp}
\end{suppfigure}

By examining in \figrefS{fig:parallel_twop_comsol} the disk's velocities $U$, $V$, and $\Omega$ versus its activity $\Pe\in[1,4]$ and the inter-disk distance $\Rgap\in ( 2, 20 ]$, we highlight the HI between two disks. At short distances, the HI enhances the swimming velocity $V$. Besides, it drives the disk to rotate at a negative $\Omega$, moving away from the its compeer---the mirror disk not shown in \figrefS{fig:parallel_twop_comsol}{A}.  $|\Omega|$ decreases with $\Rgap$ and becomes insignificant at an approximate threshold $\Rgap \approx 5$, consistent with \figrefS{fig:parallel_vel}{B}. Above this threshold, \ie $\Rgap \gtrapprox  5$, our results align with the assumed parallel configuration. Hence, the following analysis on the HI will be limited to the range $\Rgap \gtrapprox  5$.

The negative $U$ shown in \figrefS{fig:parallel_twop_comsol}{D} evidences the HI-induced attraction between disks, and the inset illustrating its magnitude $|U| \propto \Rgap^{-1}$ indicates the long-range nature of the hydrodynamic attraction. We further characterize the attractive strength by $|U/V|$ (\figrefS{fig:parallel_twop_comsol}{E}), which is observed to scale with the phoretic activity $\Pe$ as $|U/V| \propto \Pe^{5/4}$. Considering $|U| \propto \Rgap^{-1}$ and the weak dependence of $V$ on $\Rgap$ (\figrefS{fig:parallel_twop_comsol}{C}), $|U/V| \propto \Rgap^{-1}$ holds. Taken together, we infer that 
\begin{align}\label{eq:UbyV}
 |U/V| \propto \Rgap^{-1}\Pe^{\frac{5}{4}}   
\end{align}
as evidenced in \figrefS{fig:parallel_twop_comsol}{F}, which hints that the hydrodynamic attraction becomes stronger with increasing $\Pe$ and weakens with a larger inter-disk distance.

\begin{suppfigure}[tbh!]
    \centering
    \includegraphics[width=1.0\linewidth]{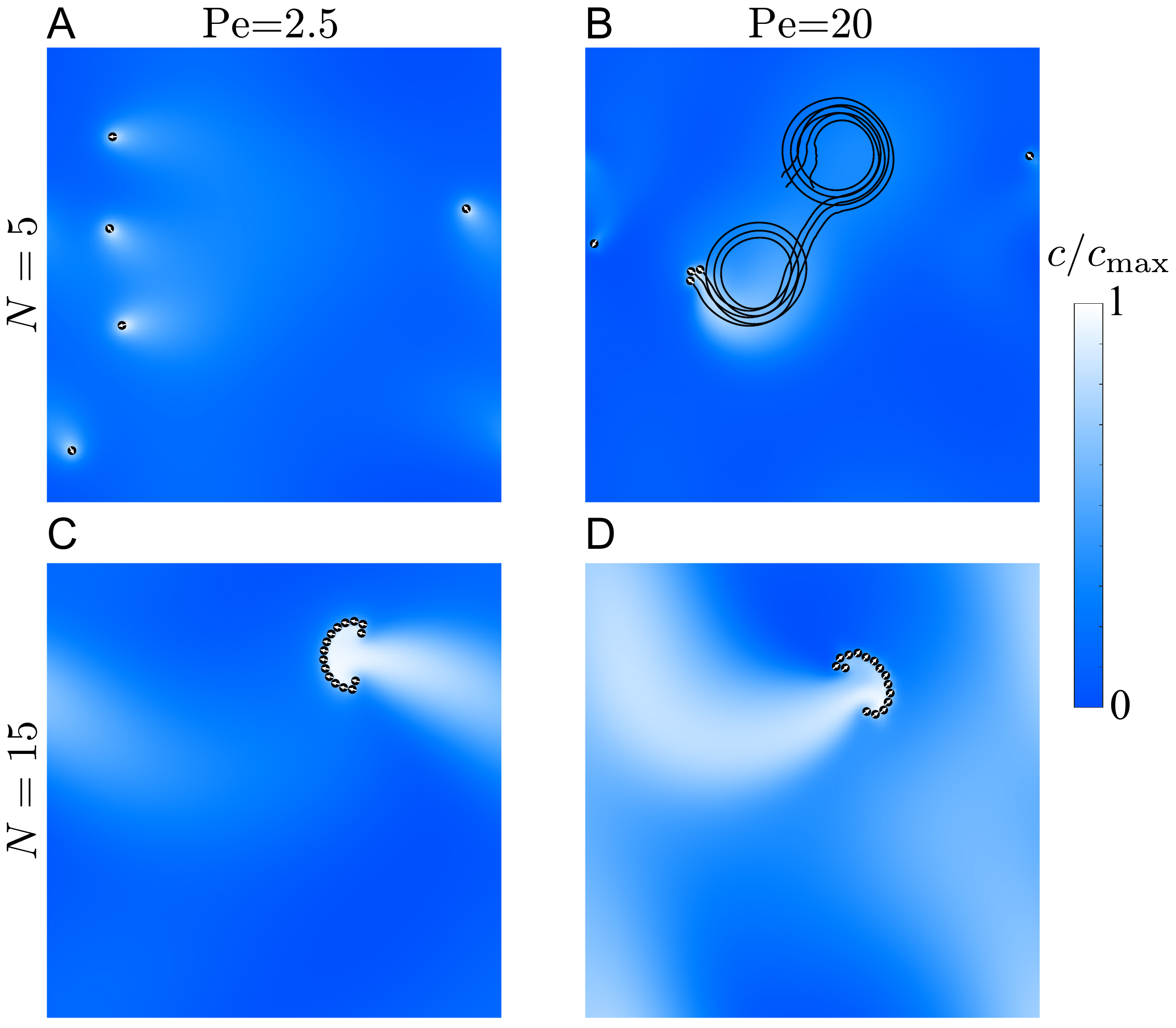}
    \caption{
    Collective behaviors of $N=5$ (top row) and $N=15$ (bottom row) disks in a periodic domain of $L=100$, when $\Pe=2.5$ (left column) and $\Pe=20$ (right column). A, freely swimming disks without forming a specific pattern. B, a three-disk chain develops, executing circular or straight trajectories intermittently, see \movRef{11}. C and D, an arc-shaped chain forms. The colormap shows the scaled concentration $c/\cmax$.
    }
    \label{fig:arc-chain}
\end{suppfigure}

We observe formation of disk chains in gas-like phases when the area fraction $\phi<0.3$, akin to the experimentally observed counterparts of chemically active droplets~\cite{thutupalli2011swarming,thutupalli2018flow} (\figrefS{fig:chain_exp}). 
Furthermore, this scenario can be divided into two categories depending on $\phi$: at an intermediate fraction, $0.05 \lessapprox \phi  \lessapprox 0.25$, disk chains form, colloid, disappear, and re-form; this phenomenon is termed dynamic chaining in the main article (Fig.~1\textbf{f} and \movRef{2}); in the dilute regime, $\phi =0.005$, a single chain swims stably as a whole. The stable chain formation is symbolized by squares in the phase diagram---Fig. 2 of the main article.

A single disk chain can be regarded as an extension of two HI-bound swimmers discussed earlier to multiple ones. Accordingly, the implication of \eqnrefS{eq:UbyV} may apply analogously: increasing the number $N$ of disks  (decreasing the inter-disk gap $\Rgap$ effectively) or the activity $\Pe$ enhances the hydrodynamic attraction, thereby facilitating chain formation.

This analogy has been  qualitatively confirmed as demonstrated below. We investigate $N\in[3,15]$ disks at $\Pe=2.5$ and $\Pe=20$ in a periodic domain sized $L=100$. The chain formation begins when $N$ reaches $5$.  When $N=5$, no chain forms at $\Pe=2.5$ (\figrefS{fig:arc-chain}{A}). The formation occurs at the higher activity $\Pe=20$. The resulting three-disk chain alternates between circular (\figrefS{fig:arc-chain}{B}) and straight motion intermittently, leaving a solute trail in its wake (\movRef{11}). When $N=15$, an arc-shaped chain emerges at both $\Pe$ values (\figrefS{fig:arc-chain}{C-D}).

It is noteworthy that similar chain formations have also been observed in other colloidal systems due to magnetic or electric dipolar interactions among colloids~\cite{schmidle2012phase,klapp2016collective}. For instance, in Ref.~\cite{schmidle2012phase}, head-to-tail chains of dipole-like colloids were reproduced using Discontinuous Molecular Dynamics simulations. A comparison between the two chain formations is given in Tab.~S\ref{tab:my-table}.

\begin{table*}[t]
\centering
\caption{
Comparison of dipole-induced chain formation as observed in Ref.~\cite{schmidle2012phase} and this study.}
\vspace{0.5em}
\label{tab:my-table}
\begin{tabular}{|l|c|c|}
\hline
  & Ref.~\cite{schmidle2012phase} & This study  \\ \hline
 Source of dipole & Magnetic or electric  & Hydrodynamic  \\ \hline
 Structure of chains & \makecell{Head-to-tail connection \\ (Fig. 5 of Ref.~\cite{schmidle2012phase})} & \makecell{Side-by-side connection \\ (\figrefS{fig:arc-chain}{C-D})} \\ \hline
 Features of dipolar interactions & \makecell{Pairwise;\\ short-ranged} & \makecell{Many-body;\\ long-ranged} \\ \hline
\end{tabular}
\end{table*}

\subsection{Two-dimensional melting}\label{sec:melting}
\begin{suppfigure}[tbh!]
    \centering
    \includegraphics[width=0.78\linewidth]{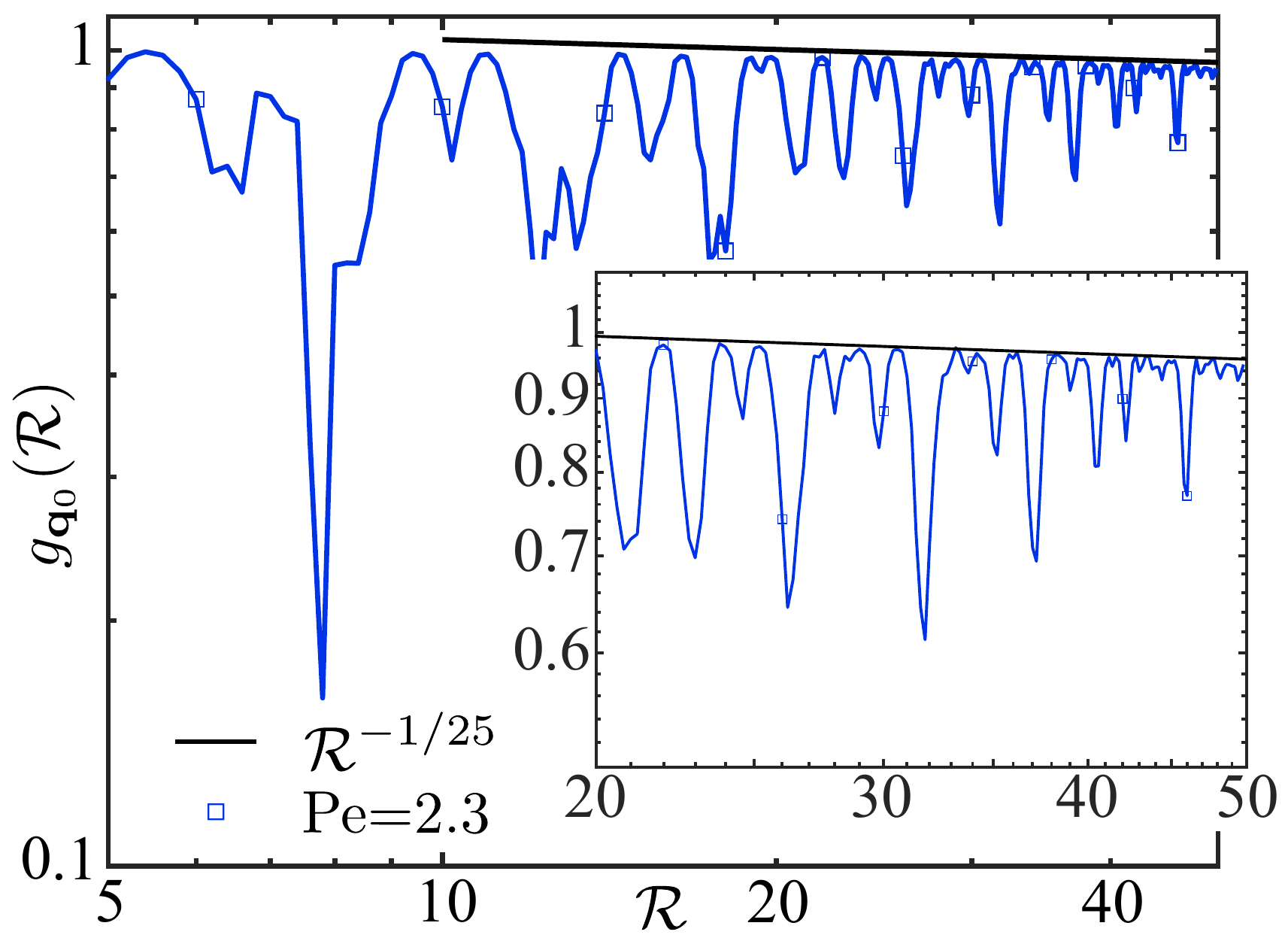}
    \caption{Translational order correlation function $\gtcfR$ for $\Pe=2.3$. The inset enlarges the scaling behavior in the large $\Rsep$ regime.
    }
    \label{fig:melting_gqr_g6r_SI}
\end{suppfigure}

\begin{suppfigure*}[tbh!]
\centering
\includegraphics[width=0.8\linewidth]{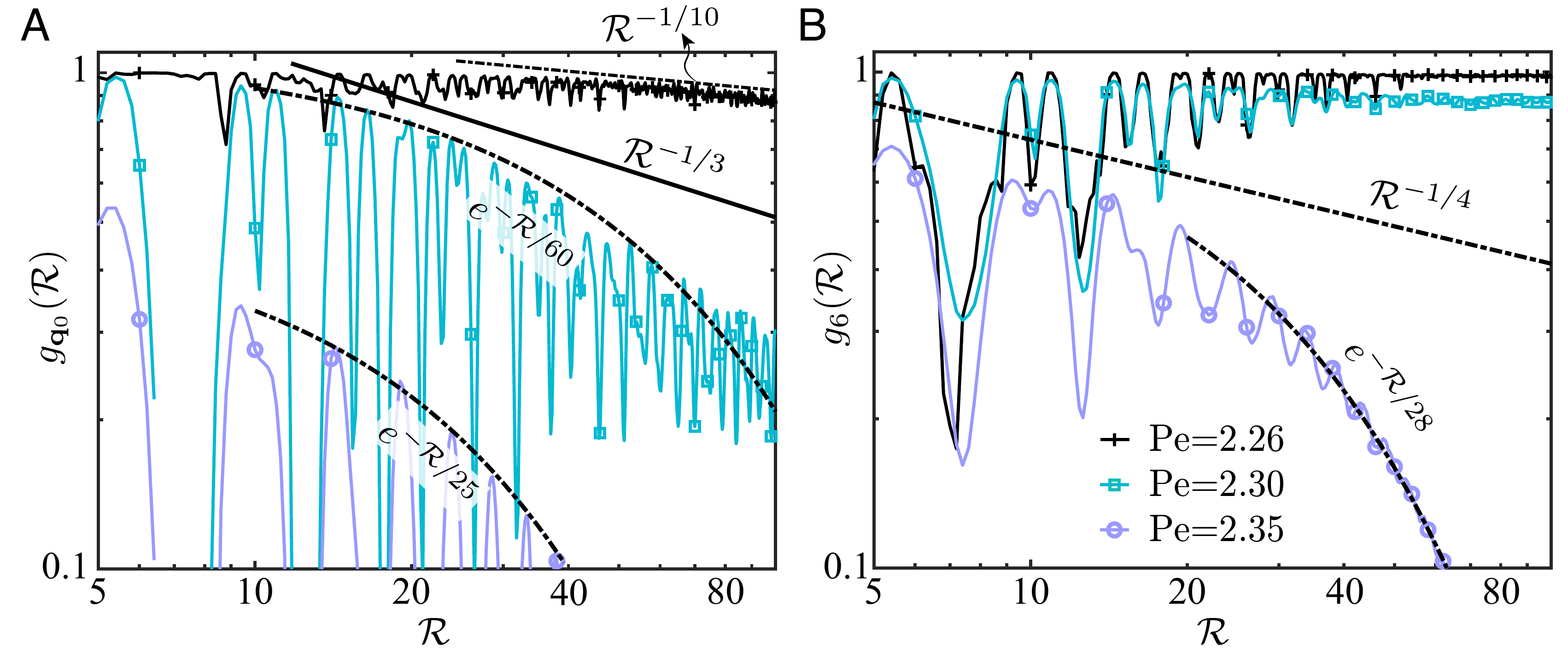}
\caption{2D melting of phoretic disks when $L=200$ as compared to previously used $L=100$ in Fig. 3 of the main article, with the area fraction $\phi=0.12$ unchanged. 
A, translational order correlation function $\gtcfR$ at $\Pe=[2.26, 2.3, 2.35]$. B, similar to A, but for the orientational order correlation function $\gocfR$.}
\label{fig:kthny_correlation_L200}
\end{suppfigure*}
In the main article (Fig. 3), we investigate the melting experienced by a disk suspension of area fraction $\phi=0.12$ as the effective temperature $\Pe$ increases. At $\Pe=2.3$, the translational order correlation function $\gtcfR \propto \Rsep^{-\etat }$ with $\eta \approx 1/25$. Here, we provide a clearer view of this scaling behavior in \figrefS{fig:melting_gqr_g6r_SI}.

Importantly, we have examined whether and how the melting dynamics depend on domain size $L$. We perform expanded simulations with a larger size $L=200$ against the $L=100$ used previously. 
We show the dependencies of the translational $\gtcfR$ and orientational $\gocfR$ correlation functions on $L$.  \figrefS{fig:kthny_correlation_L200} depicts the dependency of the translational order correlation function $\gtcfR$ and the orientational counterpart $\gocfR$ on $L$. The spatial decay of these correlation functions is found to depend on $L$ within the examined range. Nonetheless, 
the core physical picture remains unchanged. Specifically, the successive solid-to-hexatic and hexatic-to-liquid transitions reported previously persist and adhere to the KTHNY framework. Furthermore, the hexatic phase is identified at $\Pe = 2.3$, which is slightly below the previously demarcated regime $\Pe \in [2.35, 2.4]$ when $L=100$, reinforcing the solidity of our findings.

Because the fundamental findings of our study have remained consistent, we have not enlarged the domain further for a more definitive insight. In fact, the rapidly growing computational cost renders the process prohibitively resource-intensive.

\subsection{Transition to and active turbulence}\label{sec:turbulence}

\begin{suppfigure}[tbh!]
    \centering
    \includegraphics[width=1.0\linewidth]{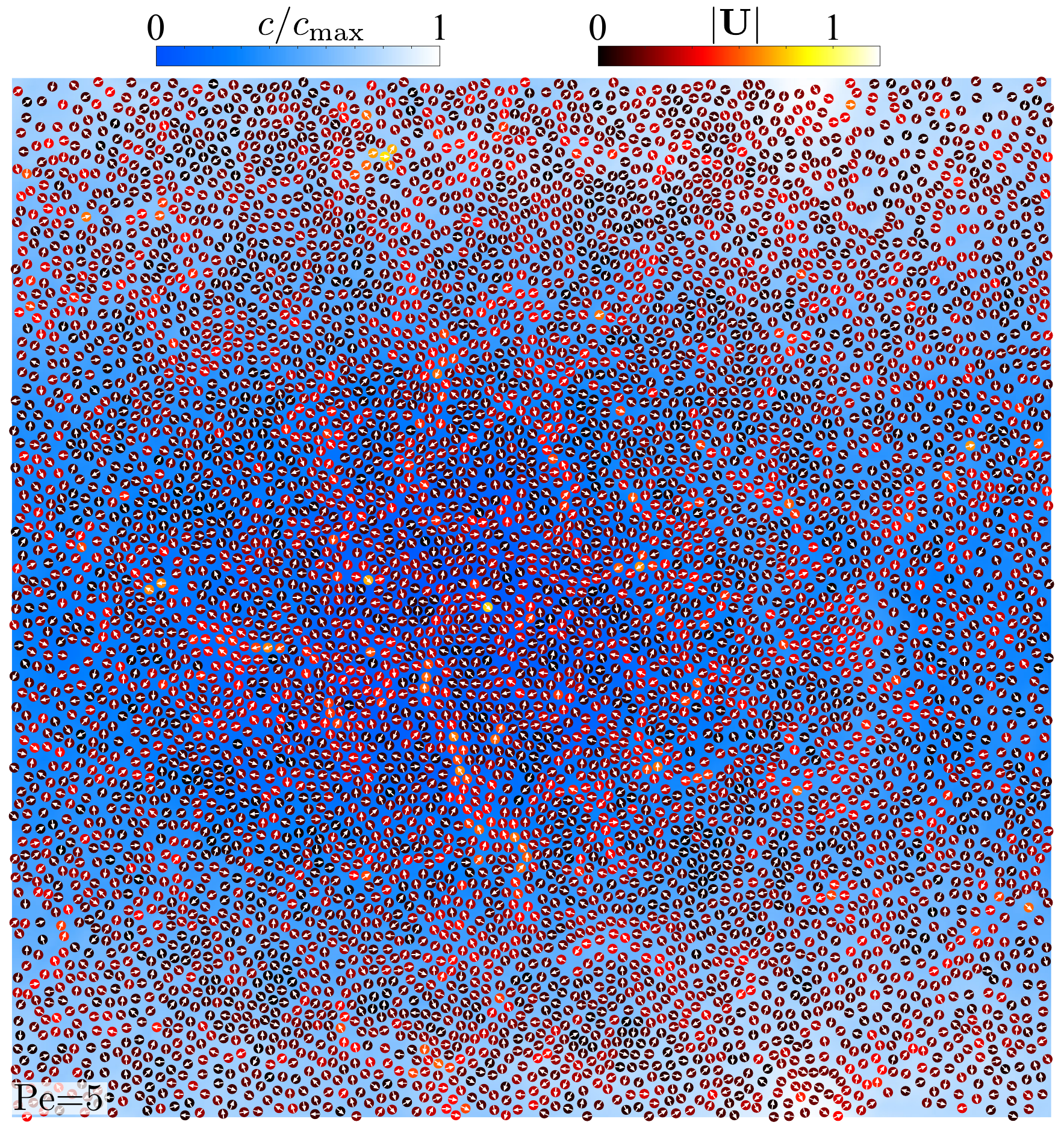}
    \caption{A snapshot of an oscillatory flow at $\Pe=5$ shows that disks concentrate in the center of the domain, contrasting with the corner concentration depicted in Fig. 4\textbf{a} of the main article. Here, $\phi=0.5$.
    }
\label{fig:wave_SI}
\end{suppfigure}

\begin{suppfigure}[tbh!]
    \centering
    \includegraphics[width=0.9\linewidth]{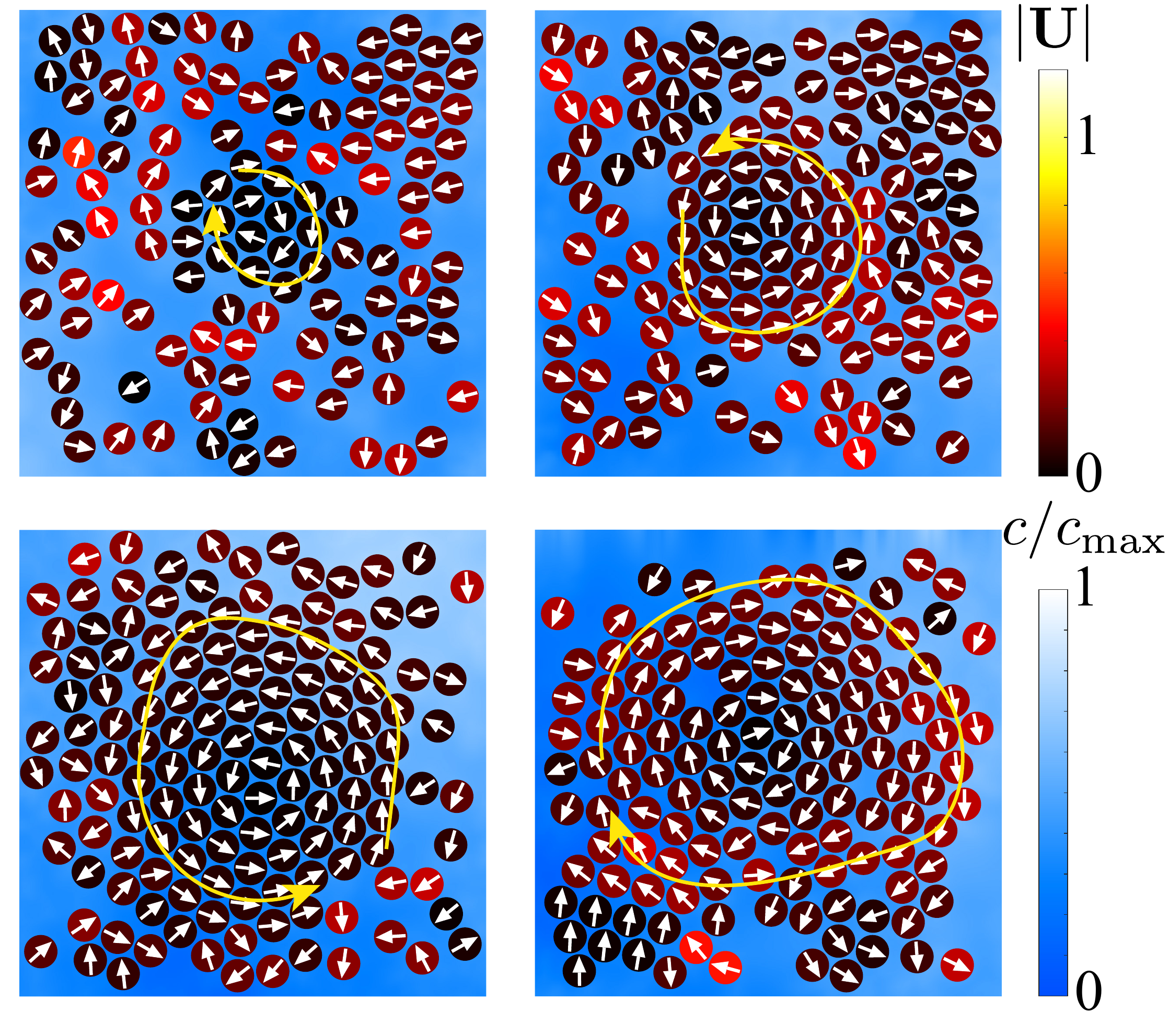}
    \caption{Lagrangian description of vortical structures formed by disks at $\Pe=20$, where $\phi=0.5$ and $L=200$. The edges of these vortexes are outlined by yellow curves.
    }
    \label{fig:Lagrangian_vortex_SI}
\end{suppfigure}

When $\Pe = 5$ and $\phi=0.5$, a self-organized oscillatory active flow of disks develops, as demonstrated in the main article; its Fig. 4\textbf{a} captures an instant when the disks concentrate in the corners of the domain. Within the oscillation, the concentrated disk region periodically relocates  between the corners and the center of the domain. For completeness, we present in \figrefS{fig:wave_SI}  the respective configuration of disk accumulation in the center.

\figrefS{fig:Lagrangian_vortex_SI} showcases the Lagrangian characterization of vortical structures formed by disks, which corroborates the Eulerian counterpart illustrated in Fig.~5\textbf{b} of the main article.

\subsubsection{Effect of finite inertia on active turbulence}\label{sec:effec_inertia}
\begin{suppfigure*}[tbh!]
\centering
\includegraphics[width=1\linewidth]{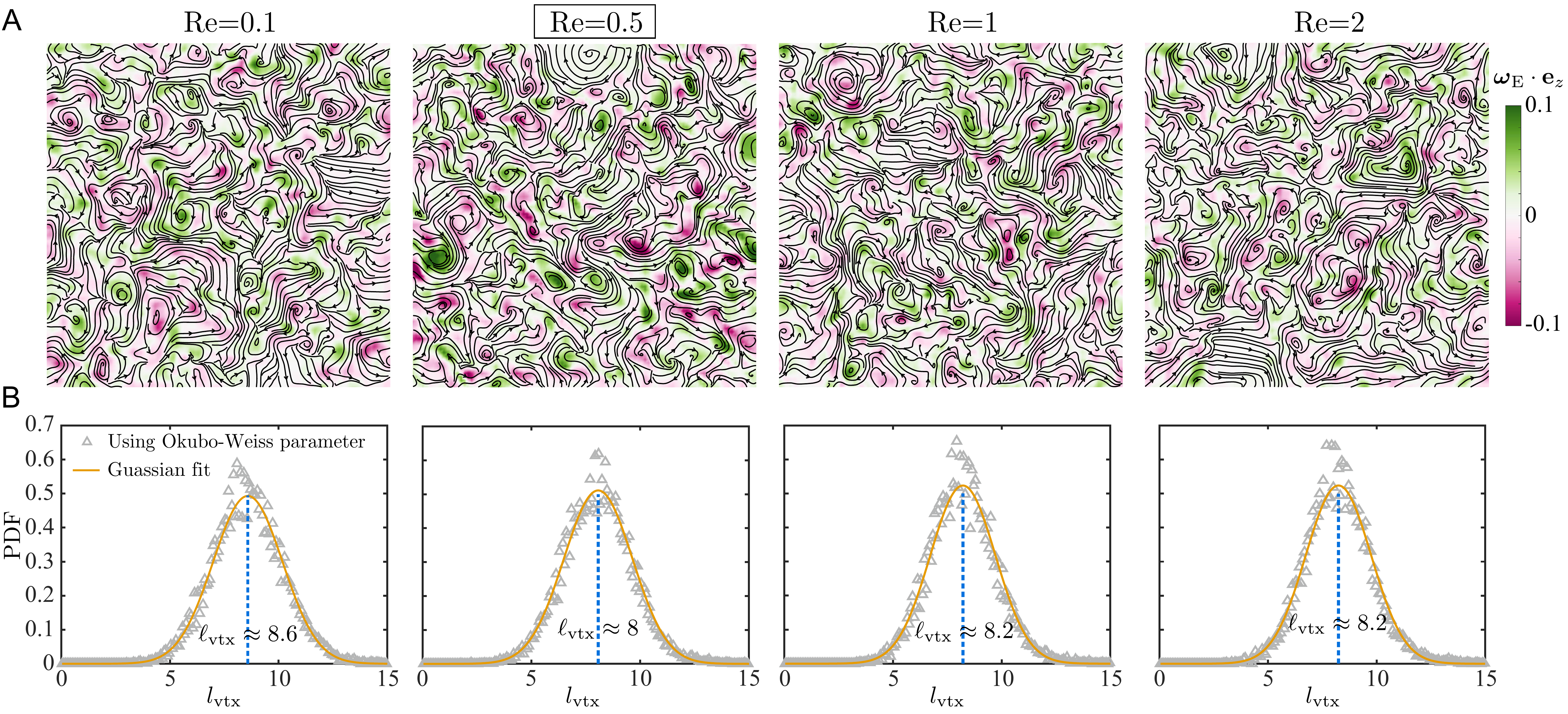}
\caption{Influence of inertia, $\Re$,  on the continuum flow field by considering the disk suspension as a continuum active fluid. In addition to our baseline case $\Re=0.5$ demonstrated in the main article, new simulations for $\Re=0.1$, $1$, and $2$ have been conducted.
Here, $\Pe=20$, $\phi=0.5$, and $L=200$. 
A, instantaneous streamlines and vorticity component $\bomegaloc \cdot \be_z$. B, size distribution of vortices identified using the Okubo-Weiss parameter.
}
\label{fig:Re_eff_vortex}
\end{suppfigure*}

\begin{suppfigure}[tbh!]
\centering
\includegraphics[width=0.95\linewidth]{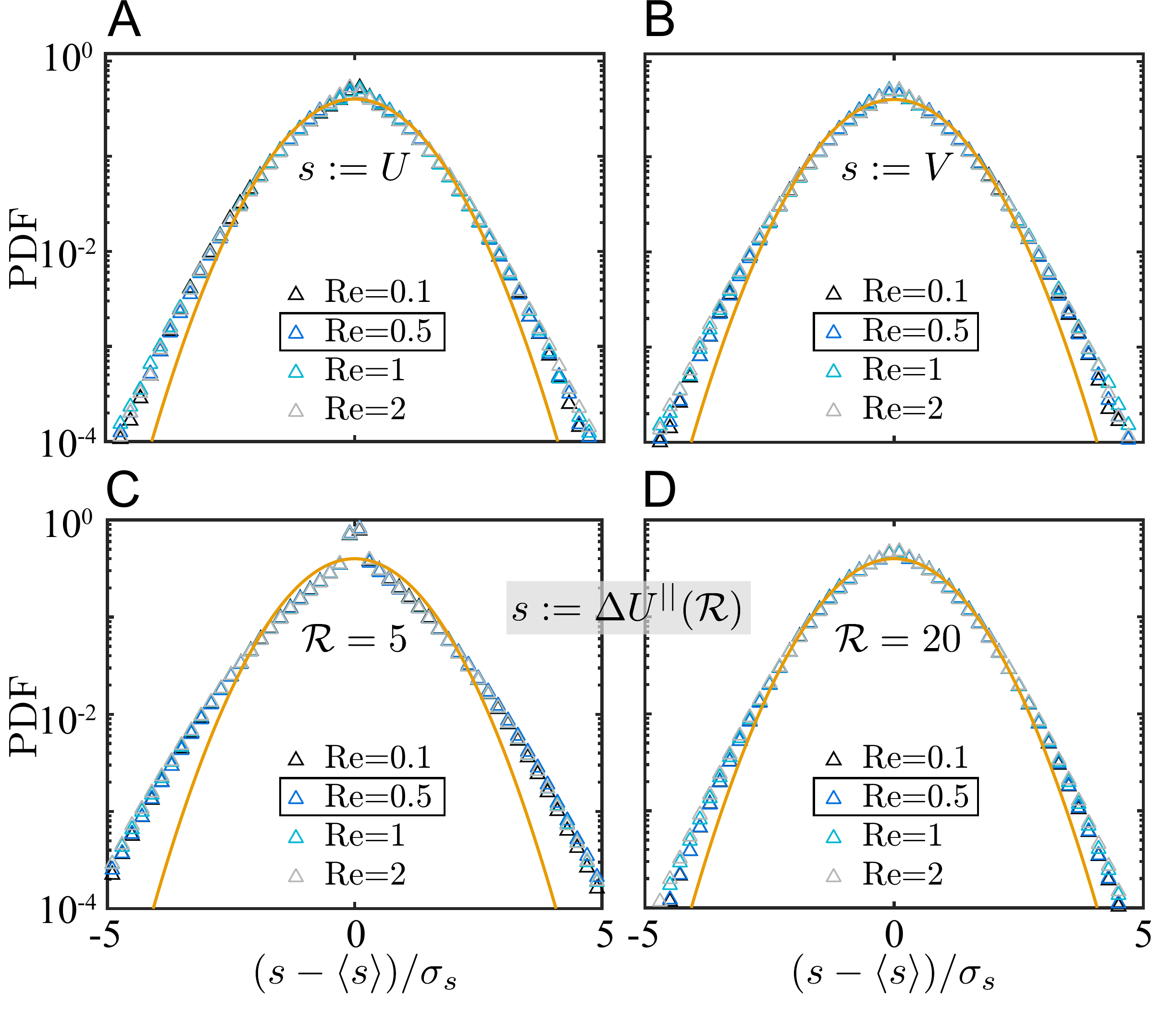}
\caption{Similar to \figrefS{fig:Re_eff_vortex}, while focusing on the statistics of disks' velocities. A and B, PDF of the disk velocity components $U$ and $V$, respectively. C,  PDF of the longitudinal component $\dUl$ of the velocity difference between two disks separated by a distance $\Rsep=5$. D, same as C, but for $\Rsep=20$.
Here, $\la s \ra$ and $\sigma_s$ represent the average and standard deviation of a random variable $s$, respectively. The curve corresponds to the unit-variance Gaussian function $1/\sqrt{2\pi}\exp(-s^2/2)$. 
}
\label{fig:Re_eff_pdf}
\end{suppfigure}

\begin{suppfigure*}[tbh!]
\centering
\includegraphics[width=0.5\linewidth]{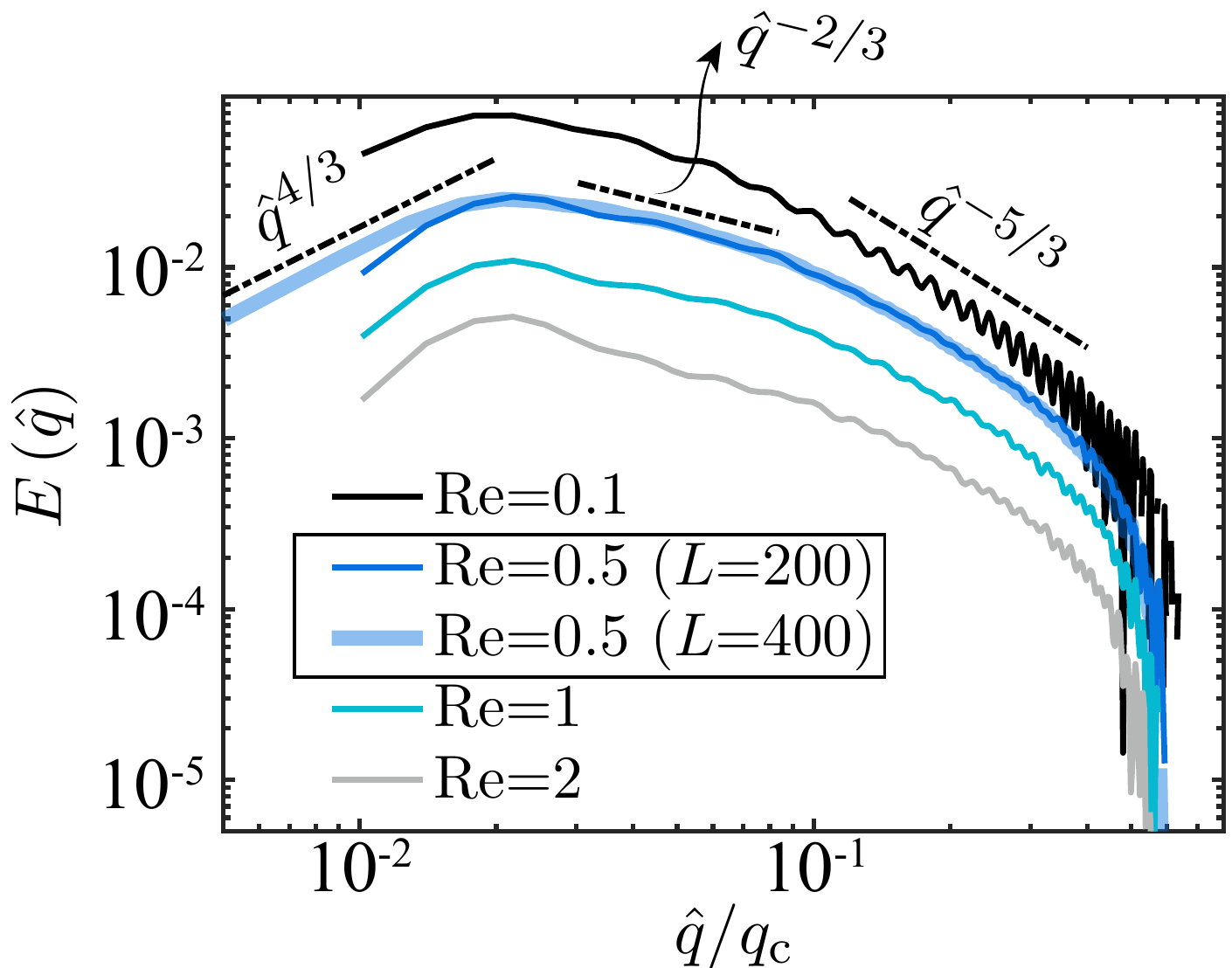}
\caption{
The effect of inertia $\Re$ on the energy spectrum $\Eqnew$ versus the modified wavenumber $\qnew$ (see Eq.~\eqref{eq:qnew}), which is vertically shifted for a better comparison. Here, $\phi=0.5$ and $\Pe=20$.
}
\label{fig:Re_eff_PSD}
\end{suppfigure*}

As indicated in \figrefS{fig:phoretic_hu}{B}, the effect of inertia $\Re=0.5$ on the propulsion of a single phoretic disk is weak.  However, this indication for single swimmers does not conclusively demonstrate that this inertia level  has negligible influence on the collective swimming dynamics.

To gain a more definitive insight, we have expanded our simulations to incorporate both smaller and larger $\Re$ values relative to our baseline setting of $\Re=0.5$. Specifically, we examine new cases with $\Re=0.1$, $\Re=1$, and $\Re=2$. To focus on the $\Re$-dependence, a domain size of $L=200$ is adopted for all the four cases.  We conduct the comparative analysis, showing the instantaneous continuum flow field and the probability density function (PDF) of vortex sizes in \figrefS{fig:Re_eff_vortex}, the PDF of the disks' velocity components and the longitudinal velocity differences in \figrefS{fig:Re_eff_pdf}, as well as the kinetic energy spectrum in \figrefS{fig:Re_eff_PSD}. 
As shown by \figrefS{fig:Re_eff_vortex} and \figrefS{fig:Re_eff_pdf}, within the studied range of $\Re \in \left[0.1, 2\right]$, we do not observe qualitative differences in the continuum flow or the statistics of disks' velocities. 
Additionally, \figrefS{fig:Re_eff_PSD} indicates that the overall trend of the energy spectrum $\Eqnew$ remains consistent across varying $\Re$. However, it is important to note that we do not assert the exact scaling exponents of $\Eqnew$ to be strictly independent of $\Re$. Such a dependency or lack thereof is, in any case, of limited consequence.

\subsubsection{Resemblance with active nematics and polar active fluids}
The recent review~\cite{alert2022active} categorizes active turbulence into two types, based on whether the contributing entities exhibit polar or nematic behaviors. Interestingly, we find that the active turbulence of IPAs, which we study, might introduce a third category. This new category demonstrates distinct characteristics from both the former categories---it shares the unique oscillatory transitional scenario with polar active fluids~\cite{giomi2012polar,alert2022active} and qualitatively exhibits the scaling behavior of the energy spectrum typical of active nematics~\cite{alert2020universal, martinez2021scaling}. 
A simple, though perhaps superficial, explanation for these observations could be attributed to the dual-state nature of an IPA. In its stable state, an IPA is stationary and thus apolar, similar to a nematic entity. However, when unstable, it behaves like a polar swimmer, akin to a Janus colloid or bacterium.

\subsection{Clustering of phoretic disks}\label{sec:clustering}
We note that clustering of active colloids has been reported before, \eg in Ref.~\cite{pohl2014dynamic}. Using Brownian dynamics simulations, 
the authors studied dilute suspensions of Janus-type phoretic colloids driven by their self-generated chemical fields, undergoing translational and rotational motion. Importantly, the study reproduces the experimentally observed dynamic clustering of Janus colloids, revealing significant clustering when the two types of motion lead to competing attractive and repulsive interactions, respectively. Here, we highlight several key differences between our work and Ref.~\cite{pohl2014dynamic} that may influence dynamic clustering:

\begin{itemize}
    \item Polarity of swimmer: Ref.~\cite{pohl2014dynamic} considers Janus-like phoretic swimmers with inherent polarity, whereas we study isotropic phoretic swimmers that spontaneously develop a swimming orientation via instability. In our case, the polarity of a swimmer can change abruptly due to the influence of surrounding swimmers.    
    \item Mechanism of attractive and repulsive interactions: In Ref.~\cite{pohl2014dynamic}, 
    the swimmer's response to the concentration gradient $\grad c$---via translational/rotational velocity---modulates the chemotaxis of swimmers. Translational motion is prescribed to induce attractive interactions, while rotational motion generates either attractive or repulsive interactions. In contrast, our study does not prescribe the kinematics or chemotactic response of our swimmers. Instead, it forms part of the solution, reflecting a Pe-dependent balance between chemical and hydrodynamic interactions. Notably, while our swimmers are chemo-repulsive, their dipolar flow produces hydrodynamic attraction as a signature of pushers.  At low Pe values, the chemical repulsion leads to the formation of a Wigner crystal. When Pe sufficiently increases, the hydrodynamic attraction overcomes the chemical repulsion, causing the swimmers to dynamically chain together. Finally, at high Pe and area fractions, dynamic clustering emerges. 
    \item Dry or wet active matter: Ref.~\cite{pohl2014dynamic} considers point swimmers without considering hydrodynamics. Instead, we account for full hydrodynamic and chemical interactions between finite-sized swimmers. 
\end{itemize}

\subsection{Reconciling two experimental observations
on camphor surfers}\label{sec:reconciling}
In the main article, we present a comprehensive simulation study that harmonizes the findings of two separate experiments on camphor surfers, which independently observed states of crystallization~\cite{soh2008dynamic} and active turbulence~\cite{bourgoin2020kolmogorovian}. The discrepancy between these observed phenomena, we suggest, is largely due to the different levels of phoretic activity, $\Pe$, presented in the two experiments. We will detail this explanation further.

Both experiments employed camphor surfers of a disk shape, albeit with different dimensions. The disk diameter is $1$ mm in Ref. \cite{soh2008dynamic}, whereas it is $5$ mm in Ref. \cite{bourgoin2020kolmogorovian}. 
Moreover, the height of disk is quite similar in both studies, specifically $500$ and $600~\mu$m respectively. Furthermore, both experiments followed the same protocol~\cite{ campbell2004arrays,smoukov2005cutting} to fabricate the disks, which should ensure their comparable levels of chemical activity. 
Besides the disk itself, another factor that can influence its behavior is the depth of the subsurface fluid, which was set at $5$ mm in Ref. \cite{soh2008dynamic} and at $10$ mm in Ref.~\cite{bourgoin2020kolmogorovian}. However, in both scenarios, these depths are considerably greater than three times the radius of the disk. As per the study~\cite{boniface2019self}, when the depth of the fluid exceeds this threshold, its influence on the disk's behavior becomes negligible.
Collectively recalling the linear dependence of $\Pe$ on the diameter, we thus deduce that the $\Pe$ of the latter experiment is approximately five  times that of the former one.
This trend indeed qualitatively aligns with our predictions (Fig. 2 of the main article), wherein lower activity leads to crystallization and higher activity induces active turbulence.

\subsection{Isotropic phoretic agents (IPAs) in an unbounded domain}\label{sec:IPAs}
While we have focused on a periodic domain in this study, it is also worth noting potential scenarios involving multiple IPAs in an unbounded domain.

Though a single unbounded IPA achieves propulsion only via instability, this is not the case for multiple IPAs. As long as $\Pe>0$, a pair of identical IPAs will consistently maintain a finite translational velocity and have zero rotational velocity. This occurs due to the disruption of their spherical or circular symmetry resulting from their mutual interaction, as highlighted in Ref.~\cite{nasouri2020exact}.

When there are more than two IPAs ($n > 2$), they will engage in translational motion, however, the disappearance of their rotational motion is not a certainty but rather contingent on their initial configuration. Despite the challenge in deciphering their behavior in a general scenario when $n > 2$, we can make a qualitative prediction for three particles ($n = 3$): their rotational motion ceases when their centers align in a straight line or are equidistant (essentially forming a regular triangle), whereas under other conditions, rotation may typically ensue.

In the unbounded case as discussed above, the particles will progressively move away from each other over time due to the inherent chemical repulsion. Ultimately, each particle will reach a degree of separation substantial enough to essentially restore an isolated state. At this juncture, the particle motion will once again depend on $\Pe$.

Unlike the unbounded case characterized by a zero area or volume fraction, the periodic domain features a finite fraction---a configuration we have adopted in our work. In such a setting, our simulations show that multiple disks initiate movement irrespective of $\Pe$. However, given a sufficiently low $\Pe$, indicative of suppressed instability, these disks will eventually self-organize into a stationary crystalline state.